\documentclass[12pt]{article}
\pdfoutput=1

\usepackage[a4paper,top=2.5cm,bottom=2.5cm,left=2cm,right=2cm]{geometry}

\usepackage[T1]{fontenc}
\usepackage[utf8]{inputenc}
\usepackage[english]{babel}

\usepackage{amsmath, amssymb, amstext, amsfonts}
\usepackage{physics}
\usepackage{mathrsfs}
\usepackage{slashed}
\usepackage{braket}

\usepackage{graphicx}
\usepackage{float}
\usepackage{subcaption}
\usepackage{adjustbox}
\usepackage{rotating}

\usepackage{multirow}
\usepackage[table]{xcolor}
\usepackage{colortbl}
\usepackage{array}
\usepackage{makecell}

\usepackage[bookmarks=false, breaklinks, colorlinks, urlcolor=black, citecolor=red, linkcolor=blue]{hyperref}

\def\bea{\begin{align}}
	\def\eea{\end{align}} 
\def\be{\begin{equation}}
	\def\ee{\end{equation}} 
\def\nn{\nonumber}
\def\Re{\text{Re}}
\def\Im{\text{Im}}

\def\tev{\ensuremath{\mathrm{Te\kern -0.1em V}}}
\def\gev{\ensuremath{\mathrm{Ge\kern -0.1em V}}}
\def\mev{\ensuremath{\mathrm{Me\kern -0.1em V}}}
\newcommand{\GeV}{\,\mathrm{GeV}}
\newcommand{\MeV}{\,\mathrm{MeV}}
\newcommand{\KeV}{\,\mathrm{KeV}}

\def\Vff#1{\ensuremath{\widetilde{\mathcal V}_{#1}}}
\newcommand{\Le}[1]{\ensuremath{\lambda_{#1}^2}}

\DeclareUnicodeCharacter{2212}{\textminus}
\newcommand{\hs}{\hspace{.4mm}}
\newcommand{\bs}{\hspace{1cm}}
\newcommand\ptwiddle[1]{\mathord{\mathop{#1}\limits^{\scriptscriptstyle(\sim)}}}

\begin{document}

\begin{flushright}
\end{flushright}

\vskip 2cm

	\begin{center}
		
		{\Large\bf Investigating non-local contributions in \\[1ex] $B_{s} \to \phi \bar{\ell} \ell$  including higher-twist effects } \\[8mm]
		{M.S.A. Alam Khan$^a$\,\footnote{Email: alam.khan1909@gmail.com},
        Rusa Mandal$^a$\,\footnote{Email: rusa.mandal@iitgn.ac.in},
        Praveen S Patil$^a$\,\footnote{Email: praveen.patil@iitgn.ac.in }
  and Ipsita Ray$^b$\,\footnote{Email: ipsitaray02@gmail.com
}}
\vskip 5pt
  {\small\em $^a$Indian Institute of Technology Gandhinagar, Department of Physics, \\ Gujarat 382355, India}
\vskip 5pt
  {\small\em $^b$Physique des Particules, Universit\'e de Montr\'eal, \\
  1375, ave Th\'er\`ese-Lavoie-Roux, Montr\'eal, QC, Canada H2V 0B3}

	\end{center}

\begin{abstract}

We analyze the impact of higher-twist three-particle  $B_s$-meson light-cone distribution amplitudes (LCDAs) on the non-local form factors for the $B_s\to \phi \bar{\ell} \ell$ transition focusing on the `charm-loop' contribution within the light-cone sum rule (LCSR) framework. To analytically continue these charm-loop contributions into the kinematically allowed region of the decay, we employ a hadronic dispersion relation that incorporates intermediate resonant states such as the $\phi,\,J/\Psi$ and $\psi(2S)$ mesons. Here, the LCSR predictions serve as inputs, supplemented by experimental data from two-body decays $B_s \to \phi ~+$ resonance states.
Our results indicate that the inclusion of twist-5 and twist-6 LCDAs enhances the non-local form factors—by approximately an order of magnitude—compared to previous estimates, due to partial disruption of cancellation among different twist contributions. This leads to a dilepton invariant mass-squared ($q^2$)-dependent correction to the Wilson coefficient $C_9$, which is higher than, but still consistent with the Standard Model prediction without the non-factorizable charm-loop corrections within uncertainties. Additionally, we update the local form factors to include contributions from higher-twist three-particle $B_s$-meson LCDAs. The phenomenological implications, particularly for the differential branching fraction and angular observables, are also discussed.

\end{abstract}
\setcounter{footnote}{0} 

\newpage
{\hypersetup{linkcolor=black}\tableofcontents}

\section{Introduction} \label{sec:intro}
Flavor-changing neutral current transitions are considered a good probe to look for physics beyond the Standard Model (SM). Within these loop-induced processes in the SM, the $b \to s \bar{\ell} \ell$ transition stands out due to its experimental significance, which is highlighted by the measurements of a multitude of decay modes such as $B \to K^{(*)} \bar{\ell} \ell$, $B_s \to \phi \bar{\ell} \ell$, and $\Lambda_b \to \Lambda \bar{\ell} \ell$, among others. While the hint of lepton flavor universality violating signatures faded away, there persist some deviations in the measurements of differential branching fractions of $B \to K \mu^+ \mu^-$~\cite{LHCb:2014cxe,CMS:2024syx}, $B_s \to \phi \mu^+ \mu^-$~\cite{LHCb:2015wdu,LHCb:2021zwz}  and in the angular observables of $B \to K^{*} \mu^+ \mu^-$ mode~\cite{LHCb:2020lmf,CMS-PAS-BPH-21-002}. Besides speculations of possible evidence of physics beyond the SM, these observables are also scrutinized with all possible relevant missing effects from the theory side. Apart from the heavy-to-light form factors, the most crucial one is known as the `charm-loop' effect, which refers to an intermediate charm loop that couples to the lepton pair with the emission of a virtual photon.

The rare semileptonic decays of the $B_{(s)}$-meson are studied in an effective theory at the $b$-quark mass scale. The effective Hamiltonian for the $b \to s \bar{\ell} \ell$ transition is written as~\cite{Bobeth:1999mk}
\begin{equation}
  \label{eq:Hamiltonian}
  \mathcal{H}_{\rm eff.}=-\frac{4G_F}{\sqrt{2}}  V_{tb}V_{ts}^*
  \Big(C_1 {\cal O}_1^c+C_2 {\cal O}_2^c+\sum_{i=3}^{10} C_i {\cal
    O}_i \Big)\,,
\end{equation}
where the contribution from the $u$-quark is doubly Cabibbo-suppressed compared to the $t$-quark part and is often neglected. The relevant operators that are dominant are
\begin{eqnarray}
  {\cal O}_7= \frac{e}{g^2} \big[\bar{s}\sigma_{\mu\nu}(m_b P_R+m_s
  P_L) b\big]F^{\mu\nu} ,~
  {\cal O}_9= \frac{e^2}{g^2}(\bar{s}\gamma_\mu  P_L
  b)\,(\bar{\ell}\gamma^\mu \ell) ,~
  {\cal O}_{10}= \frac{e^2}{g^2}(\bar{s}\gamma_\mu P_L b)\,
  (\bar{\ell}\gamma^\mu\gamma_5\ell) ,
\end{eqnarray}
where $g\,(e)$ is the strong(electromagnetic) coupling constant,
$P_{L/R}=(1\mp\gamma_5)/2$ are the left/right chiral projection
operators and $m_b\,(m_s)$ are the running $b\,(s)$ quark mass in the
$\overline{\text{MS}}$ scheme.
The expectation values of these operators with the initial and final meson states give rise to the `local' form factors whereas an important contribution of the four-quark current-current ($\mathcal{O}_{1,2}$), penguin and chromomagnetic ($\mathcal{O}_{3-6,8g}$) operators combined with the lepton-pair emission via electromagnetic interaction generates the so-called `non-local' effects as the quark-flavor transition is separated from the virtual photon emission. Among these, the dominant contribution arises from the  current-current operators with $c$-quark,
\begin{align}
  {\cal O}_1^c= (\bar{s}\gamma_\mu T^a  P_L c) (\bar{c}\gamma^\mu T^a  P_L b)~~{\rm and}~~{\cal O}_2^c= (\bar{s}\gamma_\mu  P_L c) (\bar{c}\gamma^\mu  P_L b)\,,
\end{align}
forming `charm-loops' which eventually form intermediate charmonium states ($J/\psi$, $\psi(2S)\dots$) at dilepton invariant mass squared ($q^2$) approaching the respective thresholds. The hadronic non-local matrix elements for the $B_q\to H$ type transitions, where $H$ denotes the final state meson, have been calculated in Ref.~\cite{Khodjamirian:2010vf} starting with the correlation function, 
\begin{equation}
\label{eq:H_nonlocal}
\widetilde{\mathcal{H}}^\mu (k,q) = i\int d^4x\, e^{iq\cdot
    x} Q_q\langle H(k)| T\{\bar q(x) \gamma^\mu q(x),C_1{\cal O}_1^c(0)+ C_2 {\cal O}_2^c(0) \} | B_q(k+q) \rangle\ \,,
\end{equation}
in the QCD light-cone sum rule (LCSR) approach (see also Ref. \cite{Beneke:2001at} for the calculation using the QCD factorization method and Refs. \cite{Bobeth:2010wg,Grinstein:2004vb} for the analysis at low recoil region.). Here, $Q_q$ denotes the electric charge of the quark in the electromagnetic current. First, an operator-product expansion (OPE) near the light-cone for the gluon emission from the $c$-quark loop at $q^2 \ll 4m^2_c$ is employed, giving rise to an effective non-local quark-antiquark-gluon operator. Next, the expectation value of this operator between the $B_q$-meson and the final meson state is calculated using LCSR with the $B_q$-meson light-cone distribution amplitudes (LCDAs), which serve as the non-perturbative inputs.
An improvement in the estimate of the non-local form factors was performed in Ref.~\cite{Gubernari:2020eft} with a complete set of $B_q$-meson LCDAs and updating the values of several inputs. These two analyses differ by two orders of magnitude in the estimates of non-local form factors. The parameterization of the $B_q$-meson matrix elements is decomposed in terms of the LCDAs of definite collinear twist. While the first analysis~\cite{Khodjamirian:2010vf} was performed taking into account up to twist-3 contributions, the latter one~\cite{Gubernari:2020eft} included up to twist-4 components of the $B_q$-meson LCDAs. The order difference in the estimates was mainly arising due to cancellations between the LCDAs of different twists that were missing in the earlier analysis\footnote{A recent complementary analysis~\cite{Mahajan:2024xpo} on charm-loop estimates using light-meson distribution amplitudes also reports a similar cancellation with leading twist contributions.}.

In this paper, we focus on the $B_s \to \phi \bar{\ell} \ell$ transition and revisit the estimates of both non-local and local form factors by including the twist-5 and twist-6 contributions of the three-particle $B_s$-meson LCDAs. Despite being $\mathcal{O}(1/m_{B})$ suppressed, these higher-twist LCDAs can alter the cancellation noticed in Ref.~\cite{Gubernari:2020eft} and have phenomenological implications in the observables of this mode. It has recently been pointed out that the inclusion of these higher-twist components is necessary to obtain the correct local limit of the expansion of the three-particle matrix element. In addition, it is important to note that the higher-twist two-particle LCDAs are related to the three-particle LCDAs of the same or lower twist by the QCD equations of motion (EOM)~\cite{Braun:2017liq}. Therefore, using a consistent set for both two- and three-particle LCDAs is essential and here we use such a set using only the currently known LCDAs.  A noticeable effect of these higher-twist LCDAs has been found in calculating the non-factorizable effects in the non-leptonic $B$-meson decays in Ref.~\cite{Piscopo:2023opf}. 

Also note that, apart from the higher-twist contributions which are suppressed by $\mathcal{O}(1/m_B)$, there can be additional subleading power corrections to the form factors arising from QCD factorization effects—such as power-suppressed terms from the heavy quark expansion of the hard-collinear propagator and subleading contributions from the effective weak current—which are generally computed in the Heavy Quark Effective Theory (HQET) and/or in the soft-collinear effective theory frameworks.
For the $B_{d,s} \to \pi, K$ form factors, it has been reported in Ref.~\cite{Cui:2022zwm} that among such contributions, the most significant subleading power correction arises from the two-particle twist-5 off-light-cone effect, which can alter the LCSR form factor values by approximately 25\%. Similar studies~\cite{Gao:2024vql} for $B \to K^*$ semileptonic transitions also reach a similar conclusion, emphasizing the relative impact of higher-twist effects over other subleading contributions.

For the decay of $B_s \to \phi \bar{\ell} \ell$, the parton-level transition remains the same as $B \to K^{(*)} \bar{\ell} \ell$, that is, a transition $b \to s \bar{\ell} \ell$. It shows lower observed values for the differential branching fraction in several low $q^2$-bins compared to the prediction of the theory. The different properties of $B_{u,d}$ and $B_s$-mesons are examples of the $SU(3)$ violation effects. One key parameter that enters the calculation of the form factors using $B$-meson distribution amplitude is $\lambda_B$ which is the inverse moment of the leading twist LCDA. A deviation of approximately 20\% from unity is reported in the ratio $\lambda_{B_s}/\lambda_{B}$~\cite{Khodjamirian:2020hob}. As it has already been pointed out in Ref.~\cite{Gubernari:2020eft} that the estimates of non-local form factors arising from the three-particle LCDAs are very sensitive to the inputs used, a careful look is necessary, especially at the $B_s \to \phi \bar{\ell}\ell$ predictions with suitable inputs.

The paper is organized as follows. We start in Sec.~\eqref{sec:theory} with a brief outline of the theoretical framework focusing on the LCSR calculations of the non-local (in Sec.~\eqref{sec:nonlocalFF}) form factors and local (in Sec.~\eqref{sec:localFF}) form factors including twist-5 and twist-6 contributions. Next, in Sec.~\eqref{sec:disp}, we discuss the dispersion relation techniques to access both factorizable and non-factorizable contributions in the physical $q^2$ region including the resonant states. Section~\eqref{sec:num_analysis} deals with numerical analysis and results where in Sec.~\eqref{sec:num_nonlocal}, we estimate the non-local and local form factors in the spacelike region and study their sensitivity on several input parameters, and in Sec.~\eqref{sec:fitdetails} we discuss the numerical outcomes of the fit procedure related to the analytical continuation of the charm-loop amplitudes in the timelike region. As a result, the contribution to the Wilson coefficient $C_9$ and the study of the phenomenological implications of such effects on the observables of this mode are discussed in Sec.~\eqref{sec:angobs}. Finally, we summarize the outcomes in Sec.~\eqref{sec:summary}. Several appendices are provided at the end with relevant formulae, expressions and numerical results.

\section{Theoretical framework}
\label{sec:theory}

In this section, we review the techniques related to the LCSR methods that are used to calculate both the non-local and local form factors associated with the decay $B_s \to \phi \bar{\ell} \ell$. Starting from the Hamiltonian in Eq.~\eqref{eq:Hamiltonian}, the amplitude for $B_s \to \phi \bar{\ell} \ell$ can be written as
\begin{align}
\label{eq:FullAmp}
A(B_s\to \phi \bar\ell\ell)=\frac{G_F\alpha_{\rm EM}}{\sqrt{2}\pi}V_{tb}V_{ts}^*
\bigg[&\bigg\{C_9^{\text{eff.}}\langle \phi|\bar{s}\gamma^{\mu}P_Lb| B_s\rangle 
-\frac{2 C_7^{\text{eff.}}}{q^2} \langle\phi|\bar{s} i\sigma^{\mu\nu}q_{\nu}
m_b  P_Rb|  B_s \rangle \nn \\ 
& -\frac{16\pi^2}{q^2} 
\widetilde{\mathcal{ H}}^\mu \bigg\}\, \bar{\ell}\gamma_\mu \ell + 
C^{\text{eff.}}_{10} \langle \phi | \bar{s}\gamma^{\mu}P_Lb|  B_s \rangle \,
  \bar{\ell}\gamma_\mu\gamma_5\ell \bigg],
\end{align}
where $\alpha_{\rm EM}$ is the electromagnetic coupling and the non-local matrix element $\widetilde{\mathcal{H}}^\mu$ is defined in Eq.~\eqref{eq:H_nonlocal}. The definitions of the Wilson coefficients are provided in Appendix \eqref{app:match1}. The $B_s\to \phi$ matrix elements of the (axial)vector and tensor operators can be expressed in terms of seven local form factors which depend on the momentum transfer $q^2$ between the $B_s$ (with four-momentum $p^\mu$) and $\phi$ (with four-momentum $k^\mu$) where $p^\mu\equiv (q+k)^\mu$,
\begin{align}
\langle  \phi(k) | \bar s\gamma_\mu b | B_s(p)\rangle 
&=  \epsilon_{\mu\nu\rho\sigma}\epsilon^{*\nu} p^\rho k^\sigma\,\frac{2V(q^2)}{m_{B_s}+m_{\phi}}, \label{eq:SLFF1} \\[10pt]
\langle \phi(k) | \bar{s} \gamma_\mu \gamma_5 b | B_s(p)\rangle 
&=   i \epsilon^*_\mu (m_{B_s}+m_{\phi}) A_1(q^2) 
- i (2p-q)_\mu (\epsilon^* \cdot q) \frac{A_2(q^2)}{m_{B_s}+m_{\phi}} \nonumber \\
& \quad - i q_\mu (\epsilon^* \cdot q) \frac{2m_{\phi}}{q^2}
\left[A_3(q^2)-A_0(q^2)\right], \label{eq:mat_el_V} \\[10pt]
\langle \phi(k) | \bar s \sigma_{\mu\nu} q^\nu  b | B_s(p)\rangle 
&= i\epsilon_{\mu\nu\rho\sigma} \epsilon^{*\nu} p^\rho k^\sigma \, 2 T_1(q^2), \label{eq:mat_el_T} \\[10pt]
\langle \phi(k) | \bar s \sigma_{\mu\nu} q^\nu  \gamma_5 b | B_s(p)\rangle 
&=  T_2(q^2) \left[ \epsilon^*_\mu (m_{B_s}^2-m_{\phi}^2) - (\epsilon^* \cdot q) \,(2p-q)_\mu \right] \nonumber\\
&\quad + T_3(q^2) (\epsilon^* \cdot q) \left[ q_\mu - \frac{q^2}{m_{B_s}^2-m_{\phi}^2}\, (2p-q)_\mu\right]\,\label{eq:mat_el_T1}, \\
\text{with } A_3(q^2) = \frac{m_{B_s}+m_{\phi}}{2m_{\phi}} A_1(q^2) 
&- \frac{m_{B_s}-m_{\phi}}{2m_{\phi}} A_2(q^2), ~~
A_0(0) = A_3(0) \text{ and } T_1(0) = T_2(0) . \label{eq:A30}
\end{align}
Here $\epsilon_\mu$ is the polarization vector of
the $\phi$-meson and we used the convention for the Levi-Civita tensor $\epsilon_{0123}=+1$. These local form factors are computed within two different LCSR approaches, one using light-meson distribution amplitudes~\cite{Ball:2004rg,Ball:1998kk,Bharucha:2015bzk} and another with $B$-meson LCDAs~\cite{Gubernari:2020eft}, where the later ones include both two and three particle contributions. 
The short-distance effects of $b \to s \bar{\ell} \ell$ are captured in the Wilson coefficients $C_i^{\rm eff.}$'s which are known up to next-to-next-to-leading logarithmic accuracy~\cite{Bobeth:1999mk,Altmannshofer:2008dz}.

\begin{figure}[ht]
\centering
\begin{subfigure}[b]{.25\linewidth}
\centering
\includegraphics[width=\linewidth]{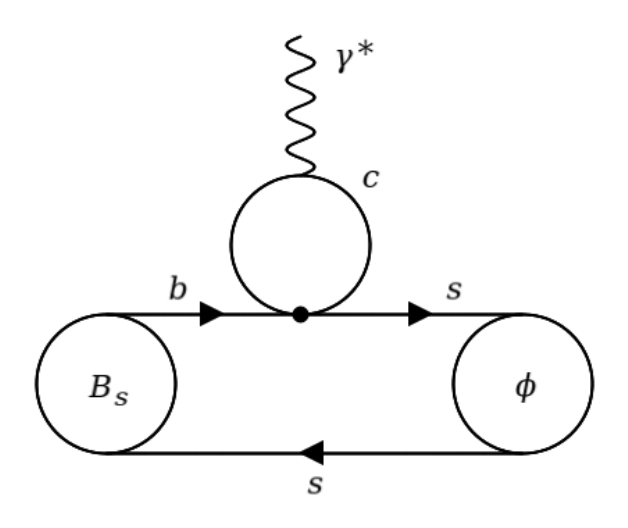}
\caption{}\label{fig:LO}
\end{subfigure} \hspace{1 cm}
\begin{subfigure}[b]{.25\linewidth}
\centering
\includegraphics[width=\linewidth]{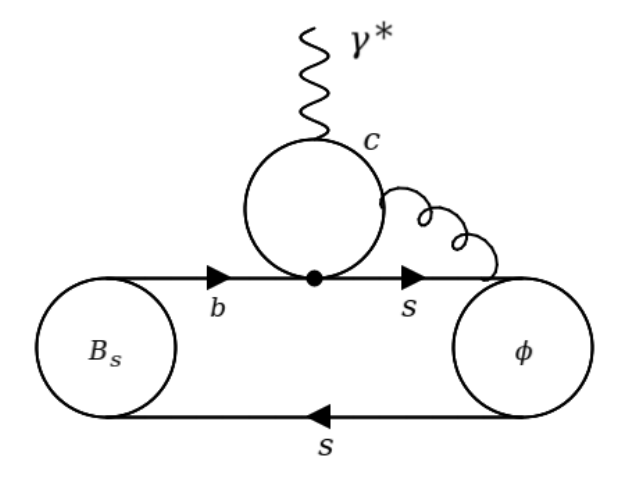}
\caption{}\label{fig:NLO1}
\end{subfigure} \hspace{1 cm}
\begin{subfigure}[b]{.25\linewidth}
\centering
\includegraphics[width=\linewidth]{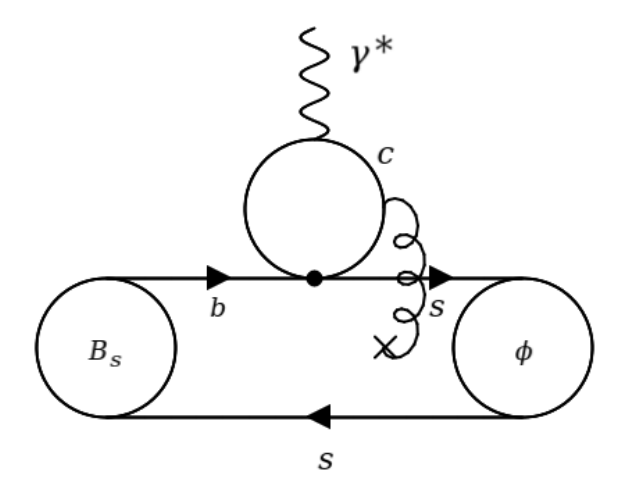}
\caption{}\label{fig:NLO2}
\end{subfigure}
\caption{Charm-loop contributions to the mode $B_s \to \phi \bar\ell \ell$ with (a)-leading-order factorizable contribution, (b)-next-to-leading-order factorizable contribution (hard gluon exchange) and (c)-non-factorizable soft-gluon correction. The dot denotes the effective four-quark vertex.}
\end{figure}

The calculation of the non-local hadronic matrix element $\widetilde{\mathcal{H}}^\mu$ proceeds in two ways relying on the OPE. The local OPE is used to compute the factorizable part arising from the leading-order diagram (Fig.~\eqref{fig:LO}) and hard gluon exchange (Fig.~\eqref{fig:NLO1}), whereas a light-cone OPE is performed to calculate the non-factorizable contribution generated due to the soft-gluon emission (Fig.~\eqref{fig:NLO2}) from the charm-loop.
In the first case, the amplitude factorizes and the non-local factorizable contributions can be included in the modifications of the Wilson coefficients multiplied by the same set of local form factors. At leading order, the local and light-cone OPE give the same result where $\widetilde{\mathcal{H}}^\mu$  can be written as
\begin{align}
16 \pi^2 \widetilde{\mathcal{H}}^\mu\vert_\text{F} = \Delta C_9(q^2) (q^\mu q^\nu - g^{\mu\nu} q^2) \langle \phi|\bar{s}\gamma_{\nu}P_Lb| B_s\rangle + \Delta C_7(q^2) 2 i m_b \langle\phi|\bar{s}\sigma^{\mu\nu}q_{\nu}
P_R b|  B_s \rangle + \cdots \,,
\label{eq:H_mu_LO}
\end{align}
where the ellipses represent the higher powers in the light-cone OPE. 
The matching coefficients $\Delta C_9$ and $\Delta C_7$ have been computed to next-to-leading order in $\mathcal{O}(\alpha_s)$ in Ref.~\cite{Asatrian:2019kbk}\footnote{Numerical values of $\Delta C_7(q^2)$ and $\Delta C_9(q^2)$ at NLO in $\mathcal{O}(\alpha_s)$ are obtained from the code provided in Ref.~\cite{Asatrian:2019kbk}.}. In order to include the soft-gluon contributions arising from the charm-loop, we make use of the light-cone OPE which was calculated for the first time in \cite{Khodjamirian:2010vf}. In this case, the $c$-quark propagator is expanded near the light-cone including the one-gluon term. The gluon field is rewritten via an exponential evolution with the non-local differential operators. The resultant matrix element giving rise to the non-factorizable contribution can then be written as
\begin{align}
\widetilde{\mathcal{H}}^\mu|_{\rm NF} = 2 Q_c\left(C_2 - \frac{C_1}{2N_c} \right) \langle \phi (k)|\widetilde{O}^{\mu} (q)| B_s (k+q)\rangle\,,
\label{eq:H_mu_NL}
\end{align}
where $N_c$ is the color factor and $\widetilde{O}^{\mu}$ is a convolution of the coefficient function $\Tilde{I}_{\mu \rho \alpha \beta}$ and non-local operator
\begin{align}
\label{eq:Onon-fac}
\widetilde{O}_{\mu}(q)= \int d\omega_2\, \Tilde{I}_{\mu \rho \alpha \beta}(q,\omega_2)\, \bar{s} \gamma^{\rho} P_L\, \delta[\omega_2 - i n_+ \cdot \mathcal{D}] G^{\alpha\beta}  b\,.
\end{align}
For the expression of the $\tilde{I}$-function we use a simplified definition provided in Ref.~\cite{Gubernari:2020eft} given by
\begin{align}
    \tilde{I}_{\mu\rho\sigma\tau} (q, \omega_2) 
        =  \int_0^1\! du \int_0^1\! dt\,
        \frac{\big( 4 t(1-t) \left( \tilde{q}_\mu \epsilon_{\rho \sigma\tau \eta} \tilde{q}^\eta 
                - 2 u \tilde{q}_\tau \epsilon_{ \mu \rho \sigma \eta} \tilde{q}^\eta
                + 2 u \tilde{q}^2 \epsilon_{\mu \rho \sigma\tau}
                \right)
                + \tilde{q}^2 \left(1 - 2u\right) \epsilon_{\mu \rho \sigma\tau}
            \big)}{64 \pi^2 ( t(1-t) \tilde{q}^2 - m_c^2)}
        \,,
\end{align}
 where $\tilde{q}^{\mu}=q^{\mu}-v^{\mu} u\hs \omega_2$ and $\tilde{q}^2\simeq q^2-u \hs \omega_2 m_b$.
Here $\omega_2$ is the $n_-$ light-cone component of the emitted gluon momentum and $v^\mu=p^\mu/m_{B_s}$ representing the four-velocity of the $B_s$-meson. In the next subsection, we compute the hadronic matrix element given in Eq.~\eqref{eq:H_mu_NL} using LCSR with the $B_s$-meson LCDAs.

\subsection{Light-cone sum rules of non-local form factors}
\label{sec:nonlocalFF}

The starting point of the LCSR calculation is the correlation function of the current and operator $\widetilde{O}_\mu$, between the $B_s$-meson and vacuum states as
\begin{equation}
\label{eq:corr_func}
{\Pi}_{\mu\nu}(q,k) = i\int d^4y\, e^{ik\cdot
    y} \langle 0| T\{\bar s(y) \gamma_\nu s(y), \widetilde{O}_\mu(q) \} | B_s(k+q) \rangle\,, 
\end{equation}
where $\bar s\gamma^\nu s$ is the $\phi$-meson interpolating current and the $B_s$-meson is taken on-shell, as a HQET state $| B_s(v)\rangle$. First, we write the hadronic dispersion relation in the variable $k^2$ for this correlation function by inserting a complete set of states with relevant quantum numbers between the interpolating current and the operator $\widetilde{O}^\mu$ as
\begin{equation}
\label{eq:Corr_disp}
{\Pi}_{\mu\nu}(q,k) =  \frac{\langle 0| \bar s \gamma_\nu s | \overline{\phi (k,\epsilon)\rangle  \langle \phi(k,\epsilon)}|\widetilde{O}_\mu(q)| B_s(k+q)\rangle }{m_\phi^2 - k^2} + \int_{s_h}^\infty ds \frac{\tilde{\rho}_{\,\mu\nu} (s,q^2)}{s-k^2}\,,
\end{equation}
where the overline denotes the average over the $\phi$ polarizations. The first matrix element in Eq.~\eqref{eq:Corr_disp} is related to the decay constant for $\phi$ as
\begin{align}
\label{eq:phi_decayC}
\langle 0| \bar s \gamma_\nu s | \phi (k,\epsilon)\rangle =  f_\phi m_\phi\, \epsilon_\nu\,.
\end{align}
The non-local matrix element can be parametrized using the similar Lorentz decomposition as of the local one but introducing a new set of form factors $\Vff{i}$, with $i=1,\,2,\,3$, reads
\begin{align}
\label{eq:nonlocal_MEdef}
\langle \phi(k)|\widetilde{O}_\mu(q)| B_s(k+q)\rangle
=& \epsilon_{\mu \alpha \beta \gamma } \epsilon^{\ast \alpha }
q^{\beta}  k^{\gamma} \Vff{1}(q^2) +i[(m_{B_s}^2-
m^2_{\phi})\epsilon^{\ast}_\mu
-(\epsilon^\ast \!\cdot q) (2k+q)_{\mu}] \Vff{2}(q^2) \nonumber \\
+& i (\epsilon^\ast\! \cdot q) \bigg[q_{\mu} - \frac{q^2}{m_{B_s}^2-
m^2_{\phi}} (2 k+q)_{\mu} \bigg] \Vff{3}(q^2))\,.
\end{align}
Substituting Eq.~\eqref{eq:phi_decayC} and Eq.~\eqref{eq:nonlocal_MEdef} in the hadronic dispersion relation in Eq.~\eqref{eq:Corr_disp}, we get
\begin{align}
{\Pi}_{\mu\nu}(q,k) &=   
   \frac{ f_\phi m_\phi }{m_\phi^2 - k^2}\bigg[\epsilon_{\mu \nu \beta \gamma } 
k^{\beta}  q^{\gamma} \Vff{1}(q^2) - i(m_{B_s}^2-
m^2_{\phi})g_ {\mu \nu} \Vff{2}(q^2) + i  q_\mu q_\nu  \Vff{23}(q^2) \bigg ] + \cdots,
\end{align}
where
\begin{equation}
\mathcal{\widetilde{V}}_{23} = \mathcal{\widetilde{V}}_{2} +\frac{ m_{\phi}^2 + q^2 - m_{B_s}^2}{m_{B_s}^2 - m_{\phi}^2}\mathcal{\widetilde{V}}_{3}\,,
\end{equation}  and only independent Lorentz structures relevant to this study are shown.
We can see that the correlation function can be decomposed in terms of scalar-valued functions $\Pi_{\mathcal{L}}(q^2,k^2)$ with independent Lorentz structures $\mathcal{L}_{\mu \nu}$ as
\begin{equation}
    \label{eq:Corr_Lortz}
    \Pi_{\mu\nu}(q, k) = \sum_{\mathcal{L}} \mathcal{L}_{\mu \nu}(q,k) \, \Pi_{\mathcal{L}}(q^2, k^2)\,,
\end{equation}
where the Lorentz structures are
\begin{equation}
\label{eq:Lorentz_struc}
\mathcal{L}_{\mu \nu}= \epsilon_{\mu\nu\beta\gamma}k^\beta q^\gamma,~g_{\mu \nu},\,q_\mu q_\nu,\,q_\mu k_\nu,\,k_\mu q_\nu,\,k_\mu k_\nu.
\end{equation}
The first three enable us to associate the $\Pi_{\mathcal{L}}$ with the form factors $\Vff{1},\,\Vff{2}$ and $\Vff{23}$, respectively.
 
The next step is to perform the OPE of the correlation function in Eq.~\eqref{eq:corr_func}. For large negative $k^2$ and $q^2\ll 4m_c^2$, the dominant contribution to the correlator comes from the region near the light-cone $y^2 \simeq 0$. We expand the strange quark propagator near the light-cone. The contraction of these $s$-quark fields gives a perturbatively calculable kernel which can be factorized from the remaining long-distance part. With the leading order term in the expansion, the correlator reads
\begin{align}
\label{eq:Corr_OPE}
\Pi_{\mu\nu}^{\rm OPE}(q,k)
= \int d\omega_2 \, \int d^4y \, \int d^4 p' \, e^{i (k - p') \cdot y}
\tilde{I}_{\mu\rho\sigma\tau} \left[ \gamma_\nu \, \frac{\slashed{p}' + m_s}{m_s^2 - p'^2} \, \gamma^\rho P_L \right]_{ab} \nn \\ 
\times\bra{0} \bar{q}_1^a (y) \delta\left[\omega_2 - i n_+ \cdot \mathcal{D}\right] G^{\sigma\tau} h^b_v(0) \ket{B_s(v)}\,,
\end{align}
where $a$,\,$b$ are spinor indices. The quantity within the square bracket represents the short-distance contribution to the correlation function. The matrix element of the non-local operator between the $B_s$-to-vacuum state is the main source of non-perturbative inputs in this light-cone OPE calculation. This can be parametrized in terms of the three-particle $B_s$-meson LCDAs defined in the HQET. The expression and detailed descriptions of the LCDAs are given in Appendix~\eqref{app:BDA}. Note that in Eq.~\eqref{eq:ME3}, there exist eight independent Lorentz structures and hence eight coefficient functions which are combinations of different twist LCDAs. We use the complete set of LCDAs including twist-5 and twist-6 components. The relevant expressions in the Exponential Model are provided in Eq.~\eqref{eq:3_LCDA_Expo}. Using the expressions given there, taking the traces, and isolating the Lorentz structures  listed in Eq. \eqref{eq:Lorentz_struc}, Eq.~\eqref{eq:Corr_OPE} can be written as
\begin{equation}
    \Pi^{\rm OPE}_{\mathcal{L}}(k^2, q^2)
        = f_{B_s} m_{B_s} \sum_n \int_0^\infty d\sigma \frac{J_{\mathcal{L}}^n(s, q^2)}{\left[k^2 - s(\sigma,q^2)\right]^n}\,,
\end{equation}
where 
\begin{equation}
    \sigma = \omega_1 / m_{B_s},\quad
    s(\sigma,q^2)=\sigma m_{B_s}^2 +\frac{m_s^2-\sigma q^2}{1-\sigma}\,.
\end{equation}
The functions $J_{\mathcal{L}}^n$ are linear combinations of the eight $B_s$-meson LCDAs, and we introduce
\begin{align}
    I_{\mathcal{L}}^n(\sigma,q^2)
    \equiv J_{\mathcal{L}}^n(\sigma,q^2)/N_\mathcal{L}=&\frac{1}{(1 - \sigma)^n} 
        \int \displaylimits_{0}^{\infty}d\omega_2 
        \int \displaylimits_{0}^{1}du 
        \int \displaylimits_{0}^{1} \left(K_1(q^2,t, u, \omega_2)\right)^{-1} dt \nn \\ &\times
        \sum_{\psi_\text{3p}}  
        \sum_{r=0}^2  
        \left(\frac{\omega_2}{m_{B_s}} \right)^r
        C^{(\mathcal{L},\psi_\text{3p})}_{n,r}(\sigma,u,t,q^2)\, 
        \psi_\text{3p} (\sigma m_{B_s}, \omega_2) 
        \, ,
\label{eq:CoeffFuncs3pt}
\end{align}
with $N_\mathcal{L}=\{+1,-i,-i\}$ respectively, for the chosen Lorentz structures (in Eq.~\eqref{eq:Lorentz_struc}), as normalization factor to ensure that the coefficients $C^{(\mathcal{L},\psi_\text{3p})}_{n,r}$ are real and matching with the convention used in \cite{Gubernari:2020eft}. Our results for these coefficients are provided in the ancillary file attached to the arXiv version of the paper. Note that here $\psi_\text{3p}\in\{\phi_3,\phi_4,\psi_4,\tilde\psi_4,\tilde\phi_5,\psi_5,\tilde\psi_5,\phi_6\}$ includes contributions up to twist-6 LCDAs and the factor 
\begin{equation}
    K_1(q^2, t, u, \omega_2) = 8\pi^2 \left[m_c^2-t(1-t)\left(q^2 - u \omega_2 m_{B_s}\right)\right]\,,
\end{equation}
is the universal part arising from the matching coefficient $\tilde{I}_{\mu\rho\sigma\tau}$ of the charm-loop calculation. 

Now using the semi-local quark-hadron duality assumption to identify the threshold $s_h$ in the hadronic dispersion relation \eqref{eq:Corr_disp} with the effective threshold of the light-cone OPE, say $s_0$, the contribution of the continuum and excited states can be eliminated. Then performing the Borel transformation in variable $k^2 \to M^2$, the LCSRs for the non-local form factors $\Vff{i}$ can be written in the following form~\cite{Gubernari:2020eft}
\begin{align}
    \Vff{i}
         = -\frac{f_{B_s} m_{B_s}}{K_2^{\Vff{i}}} &\sum_{n=1}^{\infty}\Bigg\{(-1)^{n}\int_{0}^{\sigma_0} d \sigma \;e^{(-s(\sigma,q^2)+m^2_{\phi})/M^2} \frac{1}{(n-1)!(M^2)^{n-1}}I^n_{\mathcal{L}}\nonumber\\*
        & - \Bigg[\frac{(-1)^{n-1}}{(n-1)!}e^{(-s(\sigma,q^2)+m^2_{\phi})/M^2}\sum_{j=1}^{n-1}\frac{1}{(M^2)^{n-j-1}}\frac{1}{s'}
        \left(\frac{d}{d\sigma}\frac{1}{s'}\right)^{j-1}I^n_{\mathcal{L}}\Bigg]_{\sigma=\sigma_0}\Bigg\rbrace\,,
        \label{eq:masterformula}
\end{align}
where we have introduced the following notation
\begin{eqnarray}
    &\sigma_0\equiv \sigma(s_0,q^2),\,~s'(\sigma,q^2) \equiv d s(\sigma,q^2)/d \sigma,\, \\[2ex]
    &\displaystyle\left(\frac{d}{d\sigma}\frac{1}{s'}\right)^{n} I^{k}_{\mathcal{L}}(\sigma) \equiv
    \left(\frac{d}{d\sigma}\frac{1}{s'}\left(\frac{d}{d\sigma}\frac{1}{s'}\dots I^{k}_{\mathcal{L}}(\sigma)\right)\right)\,.
\end{eqnarray}
The factors $K_2^{\Vff{i}}$ stem from the pre-factors in the hadronic side of the LCSRs, which are $$K_2^{\Vff{i}}= f_{\phi} m_{\phi},\, f_{\phi} m_{\phi} (m_{\phi}^2 - m_{B_s}^2),\,f_{\phi} m_{\phi}\,, $$
for $\Vff{1},\Vff{2}$ and $\Vff{23}$, respectively.

\subsection{Update on local form factors}
\label{sec:localFF}
In this section, we briefly comment on the updates we perform in the LCSR computation of the local form factors for $B_s \to \phi$ transition using $B_s$-meson LCDAs. We adopt a similar approach as of Ref.~\cite{Gubernari:2018wyi} where the two- and three-particle contributions of the $B_s$-meson LCDAs to the form factors are calculated. While the three-particle contributions are found to be suppressed compared to the two-particle ones, we revisit the predictions including the twist-5 and twist-6 three-particle contributions for completeness. The form factors defined in Eqs.~\eqref{eq:SLFF1} -- \eqref{eq:mat_el_T1} can be calculated starting with the correlation function similar to Eq.~\eqref{eq:corr_func} with currents  between the $B_s$-meson and the vacuum state as
\begin{align}
\label{eq:corr_funclocal}
\mathcal{F}_{\mu\nu}(q,k) = i\int d^4y\, e^{ik\cdot
    y} \langle 0| T\{\bar s(y) \gamma_\nu s(y), J_\mu^w(0) \} | B_s(k+q) \rangle, 
\end{align}
where the operator (in Eq. \eqref{eq:corr_func} )  is replaced by the much simpler weak current $J_\mu^w$ and the $\phi$ interpolating current $\bar s \gamma_\nu s$ remains the same.
It can easily be read from Eqs.~\eqref{eq:SLFF1} -- \eqref{eq:mat_el_T1} that 
\begin{equation} 
\label{eq:Jweak}
	J_\mu^w = \!
	\left\{\hspace{-1mm}
	\begin{array}{lcl}
	\displaystyle \bar s \gamma_\mu b\, : &\hspace{-5mm} & V
	\\[2ex]
	\bar s \gamma_\mu \gamma_5 b\, : &\hspace{-5mm} & A_0, A_1, A_2 \\[2ex]
	\bar s \sigma_{\mu\alpha} q^\alpha b\, : &\hspace{-5mm} & T_1 \\[2ex]
	\bar s \sigma_{\mu\alpha} \gamma_5 q^\alpha b\, : &\hspace{-5mm} & T_2,T_3\,.
	\end{array}
	\right. 
	\end{equation}
The rest of the analysis goes as in the previous section, where we first write the hadronic representation of the correlator in the variable $k^2$, separating the ground state $\phi$ from the excited and continuum states, which are encoded in the hadronic spectral density. The LCSRs for the form factors can then be obtained by
matching the hadronic representation to the OPE calculation using semi-local quark-hadron duality. For the computation of the OPE part, the correlator Eq.~\eqref{eq:corr_funclocal} is expanded at a
near light-cone separation $y^2 \simeq 0$, implying $k^2 \ll m_s^2$ and $q^2 \ll (m_b+m_s)^2$ in the momentum space. The leading term in the $s$-quark propagator generates $B_s$-to-vacuum matrix element $\left\langle 0\left|\bar{s}^a(y) h_v^b(0)\right| B_s(q+k)\right\rangle$ which can be approximated in terms of the two-particle LCDAs (see Eq.~\eqref{eq:BLCDAs2pt}) whereas the gluon emission gives rise to the following matrix element $\bra{0} \bar{s}^a(y)
G_{\sigma\tau} h_v^b(0) \ket{B_s(v)}$ parameterized in terms of the three-particle LCDAs (see Eq.~\eqref{eq:MEL3}). We do not repeat all the expressions here and refer the reader to Eq.(2.18) of Ref.~\cite{Gubernari:2018wyi}. The only difference in our case arises in the function $I_n^{(F,\,3p)}$ (in Eq. (2.21) of \cite{Gubernari:2018wyi}) where the sum is taken over all the three-particle LCDAs with increasing twists, namely $\psi_\text{3p}\in\{\phi_3,\phi_4,\psi_4,\tilde\psi_4,\tilde\phi_5,\psi_5,\tilde\psi_5,\phi_6\}$. Note that the $I_n^{(F,\,3p)}$ function contains $1/\omega_2$ factor within the integrand which is compensated by the $w_2$ dependence in the numerator for all the $\psi_\text{3p}$ except $\tilde{\phi}_5$ and $\phi_6$ written in the Exponential Model. However, interestingly such diverging contributions arising from this $1/\omega_2$ dependence in the integration are canceled between different terms, ensuring the stability of the LCSRs.

\section{Dispersion relation for non-local amplitudes}
\label{sec:disp}

In this section, we combine the total non-local effect in the decay mode $B_s\rightarrow\phi \bar{\ell} \ell$ obtained using the light-cone OPE with the factorizable correction in Eq.~\eqref{eq:H_mu_LO} and the non-factorizable soft gluon correction in Eq.~\eqref{eq:H_mu_NL} as
\begin{equation}
    \widetilde{\mathcal{H}}_\mu(q^2)= \widetilde{\mathcal{H}}_\mu\vert_\text{F}+\widetilde{\mathcal{H}}_\mu\vert_\text{NF}\,.
\end{equation}
The higher-order corrections in the light-cone expansion are assumed to be small and therefore neglected in this analysis.\par 
At this point, it is also convenient to rewrite the definitions of the form factors in the transversity basis which is the widely adopted convention and directly relates the polarization of the final state meson and the dilepton pair. This is achieved by reorganizing the combination of Lorentz structures present in the definitions of local and non-local matrix elements (Eqs.~\eqref{eq:SLFF1} -- \eqref{eq:mat_el_T1}, and Eq.~\eqref{eq:nonlocal_MEdef}). The relationship between the traditional and transversity bases of form factors is presented in Appendix~\eqref{app:had}.\par
In the transversity basis, $\widetilde{\mathcal{H}}_\mu(q^2)$ in Eq.~\eqref{eq:H_nonlocal} can be decomposed as
\begin{equation}
\label{eq:Hmu_NLlam}
    \widetilde{\mathcal{H}}_\mu
        = m_{B_s}^2\, \epsilon^{*\alpha}\left[
        \mathcal{L}_{\alpha\mu}^{\perp} \widetilde{\mathcal{H}}_\perp - \mathcal{L}_{\alpha\mu}^{\|} \widetilde{\mathcal{H}}_\|  - 
        \mathcal{L}_{\alpha\mu}^{0} \widetilde{\mathcal{H}}_0\right]\,,
\end{equation}
where $\mathcal{L}_{\alpha\mu}^\lambda$ are Lorentz structures given in Eq.~\eqref{eq:def:Lor} of  Appendix~\eqref{app:had}. The non-local transversity amplitudes $ \widetilde{\mathcal{H}}_\lambda$, for $\lambda=\{\perp,\parallel,0\}$, read 
\begin{align}
    \label{eq:NLFF_full}
   \widetilde{\mathcal{H}}_\lambda &=
    -\frac{1}{ 16 \pi^2 }
    \left(
        \frac{q^2}{2m_{B_s}^2}
        \Delta C_9\, \mathcal{F}_\lambda+\frac{m_b}{m_{B_s}}\Delta C_7\,
        \mathcal{F}_{\lambda,T}
    \right)
    +
    2 \,Q_c\, \left(C_2 - \frac{C_1}{2N_c} \right)
    \Vff{\lambda}\,,\nonumber \\
    &\equiv\mathcal{H}_\lambda-\frac{1}{ 16 \pi^2 }\frac{m_b}{m_{B_s}}\Delta C_7\,        \mathcal{F}_{\lambda,T}    \,, 
\end{align}
where $\mathcal{F}_\lambda\, (\mathcal{F}_{\lambda,T})$ are the local vector (tensor) form factors in transversity basis (see Eq.~\eqref{eq:ff_convert} for the relation to traditional form factors). Note that in the second line of the above equation, we have separated the term proportional to $\Delta C_7$ which corresponds to the real photon emission and hence contributes dominantly to the $B_s\rightarrow \phi \gamma$ process. The rest of the effects are collected in the non-local transversity amplitude $\mathcal{H}_\lambda(q^2)$ which will be the relevant piece for the discussion in the rest of the article. 
Our next purpose is to analytically continue $\mathcal{H}_\lambda$ in the timelike region via the use of dispersion relation in the variable $q^2$. We insert a complete set of $(q\bar q)$ vector meson bound states in between the quark electromagnetic current $\bar q \gamma_\mu q$ and the four-quark operator in the correlation function~\eqref{eq:H_nonlocal}, and separate the lowest-lying states such as $\phi$ for the $s$-flavor case and $J/\psi$ and $\psi(2S)$ in the case of $c$-flavor using the following amplitude definitions \cite{Khodjamirian:2010vf},
\begin{align}
\langle 0 | \bar q \gamma^\mu q | V (q) \rangle &= \epsilon_V^\mu m_V f_V\,,\label{eq:f_def_decay} \\
\langle \phi (k) V (q) |\left( C_1 O_1 + C_2 O_2 \right) | B (k+q) \rangle &=i \sqrt{2}\epsilon_\alpha \epsilon_{V\beta}\Big\lbrace i \epsilon^{\alpha\beta \rho\tau}k_\rho q_\tau \frac{A^\perp_V}{\tilde{\lambda}_{m_V^2}^{1/2}}+ g^{\alpha\beta} \frac{A^\parallel_V}{2}\nonumber\\&\hspace{-4cm}+ \frac{1}{\tilde{\lambda}_{m_V^2}}  (k+q)^\alpha(k+q)^\beta  \times \big[2\sqrt{2} m_\phi m_V  A^0_{V} -(m_{B_s}^2-m_V^2-m_\phi^2)A^\parallel_V\big] \Big\rbrace\,,\label{eq:res_amp_def}
\end{align}
where\footnote{ Note an extra $\sqrt{2}$ pre-factor in front of $A^0_V$ in Eq.~\eqref{eq:res_amp_def} which is missing in Ref.~\cite{Khodjamirian:2010vf}.} we have used the K\"all\'en function $\tilde{\lambda}_{X^2}\equiv\lambda(m_{B_s}^2,m_{\phi}^2,X^2)$ = $m_{B_s}^4+m_{\phi}^4+X^4-2(m_{B_s}^2 m_{\phi}^2+ m_{B_s}^2 X^2+m_{\phi}^2 X^2)$. 
This allows us to write the resonance contributions in terms of the hadronic two body $B_s \to \phi V$  decay amplitudes $A_V^\lambda$ relying on Breit-Wigner ansatz which takes into account the finite width effects of the resonances. Due to the dependency of the form factors $\mathcal{F}_\lambda$ on variable $q^2$ (see Eq.~\eqref{eq:ff_convert}) in the expression of non-local amplitudes  $\mathcal{H}_\lambda$ (in Eq.~\eqref{eq:NLFF_full}), we use double-subtracted (at $q_0^2$) dispersion relation for $\mathcal{H}_\lambda$ as  
\begin{align}
 \mathcal{H}_{\perp,\|}(q^2)&= \mathcal{H}_{\perp,\|} (q_0^2) + (q^2-q_0^2)\frac{d}{d\hs q^2}  \mathcal{H}_{\perp,\|} (q^2)\bigg\vert_{q^2=q_0^2} + (q^2-q_0^2)^2 \times \nn \\[2ex]  \Bigg\lbrace & \sum_{V=\phi,\,J/\psi,\,\psi(2S)} \frac{\kappa_V
f_V m_V |A_{V}^{\perp,\|}| e^{-i (\varphi_V^{\perp,\|}-\varphi_V^0)} }{m_{B_s}^3(m_V^2-q_0^2)^2(m_V^2-q^2-
im_V\Gamma^{\text{tot.}}_V)} + \int_{s_0^h}^{\infty}ds\hs 
\frac{\rho_{\perp,\|}(s)}{(s-q_0^2)^2(s-q^2-i\epsilon)} \Bigg\rbrace \,, \label{eq:disppsi}\\
 \mathcal{H}_{0}(q^2)&= \mathcal{H}_{0} (q_0^2) + (q^2-q_0^2)\frac{d}{d\hs q^2}  \mathcal{H}_{0} (q^2)\bigg\vert_{q^2=q_0^2} + (q^2-q_0^2)^2 \times \nn \\[2ex]  \Bigg\lbrace & \sum_{V=\phi,\,J/\psi,\,\psi(2S)} \frac{\kappa_V
f_V m_V^2 |A_{V}^{0}| e^{i \varphi_V^{0}} }{m_{B_s}^4(m_V^2-q_0^2)^2(m_V^2-q^2-
im_V\Gamma^{\text{tot.}}_V)} + \int_{s_0^h}^{\infty}ds\hs 
\frac{\rho_{0}(s)}{(s-q_0^2)^2(s-q^2-i\epsilon)} \Bigg\rbrace\,,
\label{eq:disppsi0}
\end{align}
where $\kappa_V$  is $-1/3$ for the $\phi$-meson and $+2/3$ for the charmonium states which correspond to the charge of the electromagnetic current associated with these quarkonium states.
Other higher resonances above the $D\bar D$ threshold i.e., with $q^2> 4m_D^2$ are treated as part of the continuum and their contributions can be modeled either using a single pole approximation used in Ref.~\cite{Khodjamirian:2010vf} or a fit function linear in $q^2$ used in Ref.~\cite{Khodjamirian:2012rm}. Recent studies~\cite{Mutke:2024tww,Gopal:2024mgb} have shown that apart from the unitarity branch cut starting from $q^2=(m_{B_s} + m_\phi)^2$, the multiparticle states arising from non-local operators exhibit anomalous cuts after $4m_D^2$ in the complex $q^2$ plane. We refrain from delving into the detailed structure of such anomalous cuts and use the simple linear model in $q^2$ in our analysis which uses the following parametrization 
\begin{equation}
\int_{4m_D^2}^{\infty} ds
\frac{\rho_{\lambda} (s)}{(s-q_0^2)^2(s-q^2-i\epsilon)}= a_{\lambda} + b_{\lambda} \,  \frac{q^2}{4 m_D^2}\,.
\label{eq:model_larges}
\end{equation}
Here $a_\lambda$ and $b_\lambda$ are unknown complex parameters to be determined from fitting to LCSR predictions which we discuss in Sec.~\eqref{sec:non-local FF}. No significant numerical effects are observed if we use the single pole approximation instead of Eq.~\eqref{eq:model_larges}. Several alternative parameterizations for non-local form factors, inspired by the $z$-parametrization and the analyticity properties of non-local amplitudes, can be found in Refs. \cite{Bobeth:2017vxj,Gubernari:2020eft}. These approaches also incorporate experimental data on two-body $B_{(s)}\to V K^* (\phi)$ transversity amplitudes, which dominate the strength of non-local amplitudes near the resonance regions. Since our primary focus is on the low-$q^2$ region, we do not anticipate a significant impact of the $z$-parametrization on extrapolation compared to the dispersion relation analysis used in this paper. However, it would be interesting to explore how the enhancement of LCSR estimates for the non-local form factors influences the global dispersive analysis of $b\to s$ transitions recently performed in Ref. \cite{Gubernari:2023puw}. This study, which combines $B \to K^{(*)}$ and $B_s \to \phi$ modes, employs dispersive bounds to constrain the $z$-expansion coefficients, thereby mitigating systematic uncertainties arising from the truncation of the $z$-expansion.


Note that in Eqs.~\eqref{eq:disppsi} and \eqref{eq:disppsi0} we have also explicitly separated the phase $\varphi_V^\lambda$ in the polarization amplitudes, for each resonance,  which are defined relative to the longitudinal component $\varphi_V^0$. Phases $\varphi_V^\perp$ and $\varphi_V^{||}$ are obtained from the LHCb results in Refs.~\cite{LHCb:2023exl,LHCb:2023sim,LHCb:2016tuh} and $\varphi_V^0$ for the three resonances are considered as free parameters in the fit. On the other hand, the amplitudes $A_{V}^\lambda$ are obtained from the polarization fraction measurements of the decays namely,
$f_\lambda \equiv|A_{V}^\lambda|^2/\left(\sum_\lambda \vert A^\lambda_{V}\vert^2\right) $~\cite{LHCb:2023exl,LHCb:2023sim,LHCb:2016tuh} where the decay width of the resonances are used for the normalization of the sum of all three polarization amplitudes via
\begin{equation}
    \Gamma(B_s\rightarrow \phi\hs V)=\frac{\tilde{\lambda}_{m_V^2}^{1/2}}{16\pi m_{B_s}^3 } \left(\frac{4 G_F}{\sqrt{2}}\right)^2 \vert V_{c b}V_{c s}^*\vert^2 \sum_{\lambda=\perp,\parallel,0} \vert A^\lambda_{V}\vert^2 \,.
    \label{eq:decay_resonance}
\end{equation}

\section{Numerical analysis and results}
\label{sec:num_analysis}

\begin{table}[h]
    \centering
    \renewcommand*{\arraystretch}{1.1}
    \begin{tabular}{||c|c|c c||}\hline\hline
         Parameter description&Symbol& Value &Ref.  \\ \hline\hline
        $B_s$-meson mass&   $m_{B_s}$& $5366.93\pm0.10\MeV$&  \cite{ParticleDataGroup:2022pth} \\
         \hline Decay constant of $B_s$-meson &
         $f_{B_s}$& $230.3\pm1.3\MeV$ &  \cite{FlavourLatticeAveragingGroupFLAG:2021npn}   \\
         \hline
        \multirow{3}{*}{$B_s$-meson LCDA parameters}&   $\lambda_{B_s}$& $480\pm92\MeV$&  \cite{Mandal:2024pwz}  \\
         \cline{2-4}
      &   $\lambda_{E}^2$& $0.03\pm 0.02\GeV^2$ &  \cite{Nishikawa:2011qk}   \\
         \cline{2-4}
         & $\lambda_{H}^2$& $0.06\pm 0.03\GeV^2$ &  \cite{Nishikawa:2011qk}    \\
         \hline $\phi$-meson mass   & $m_\phi$& $1019.461\pm0.016\MeV$ &\cite{ParticleDataGroup:2022pth} \\ \hline
        Effective threshold & $s_0$  &$[2.1,2.4]\GeV^2$ & \cite{Gubernari:2020eft} \\ \hline
          Charm quark pole mass &$m_c\equiv  M_c$& $1.67\pm0.07\GeV$ &  \cite{ParticleDataGroup:2022pth}  \\
         \hline Strange quark mass&
           $m_s(2\GeV)$& $93.5\pm 0.8\MeV$& \cite{ParticleDataGroup:2022pth}\\ \hline\hline
    \end{tabular}
    \caption{Parameters used in the estimates of the local and non-local form factors.}
    \label{tab:inputs}
\end{table}

In this section, we present the results of the numerical analysis based on the theoretical framework described in the previous sections. This is divided into several subsections as follows.

\subsection{\texorpdfstring{}{} Form factor \texorpdfstring{estimates}{} in spacelike region}
\label{sec:num_nonlocal}

First, we analyze the impact of incorporating higher-twist contributions on the LCSR predictions for both the non-local and local form factors. Next, we examine the sensitivity of these predictions to the key non-perturbative parameters, $\lambda_E^2$ and $\lambda_H^2$, which represent the local matrix elements associated with the quark-gluon three-body components in the $B_s$-meson LCDAs. 

\subsubsection{Higher twist corrections}

Based on the formalism discussed in Sec.~\eqref{sec:nonlocalFF}, we calculate the non-local form factors using the LCSRs given in Eq.~\eqref{eq:masterformula}. The inputs entering the expression are listed in Table~\eqref{tab:inputs}. Apart from the standard constants and mass parameters, the intrinsic parameters for the LCSR are the effective threshold $s_0$ and the Borel window $M^2$ which is taken as $0.75\,\text{GeV}^2 < M^2 < 1.25\,\text{GeV}^2$ \cite{Gubernari:2020eft}. We have verified that the inclusion of higher-twist effects does not alter the window, and the sum rule dependence on $M^2$ remains mild. For the $B_s$-meson LCDAs, we consider the widely used Exponential Model in our analysis. The expressions are quoted in Appendix \eqref{app:BDA}. 

\begin{table}[H]
\centering
\renewcommand*{\arraystretch}{1.2}
\begin{adjustbox}{max width=\textwidth}
\begin{tabular}{||c||c||c|c|c||c|c|c||c|c|c||}
\hline\hline
\multirow{2}{*}{$\lambda$} & $q^2$  & \multicolumn{3}{c||}{$\mathcal{F}_{\lambda}$} & \multicolumn{3}{c||}{$\mathcal{F}_{\lambda,T}$} & \multicolumn{3}{c||}{$\Vff{\lambda} \times 10^7$} \\ 
\cline{3-11}
 & $\GeV^2$ & Twist-4 & Twist-6 & Ref.\cite{Gubernari:2020eft} & Twist-4 & Twist-6 & Ref.\cite{Gubernari:2020eft} & Twist-4 & Twist-6 & Ref.\cite{Gubernari:2020eft}            \\ 
\hline\hline
\multirow{5}{*}{0} 
 & -9  & $0.350\pm0.072$ & $0.350\pm0.072$ & -- & $-0.116\pm0.024$ & $-0.116\pm0.024$ & -- & $1.709\pm0.812$ & $7.438\pm4.520$ & -- \\ 
\cline{2-11}
 & -7  & $0.357\pm0.072$ & $0.360\pm0.074$ & $0.315\pm0.180$ & $-0.093\pm0.019$ & $-0.094\pm0.019$ & $-0.088\pm0.035$ & $1.235\pm0.597$ & $6.250\pm3.857$ & $-1.7\pm0.7$  \\ 
\cline{2-11}
 & -5  & $0.368\pm0.074$ & $0.371\pm0.075$ & $0.330\pm0.188$ & $-0.069\pm0.014$ & $-0.069\pm0.015$ & $-0.066\pm0.035$ & $0.792\pm0.396$ & $4.862\pm3.052$ & $-1.1\pm0.5$  \\ 
\cline{2-11}
 & -3  & $0.379\pm0.076$ & $0.382\pm0.077$ & $0.341\pm0.194$ & $-0.043\pm0.009$ & $-0.043\pm0.009$ & $-0.041\pm0.021$ & $0.402\pm0.217$ & $3.209\pm2.053$ & $-0.58\pm0.27$ \\ 
\cline{2-11}
 & -1  &$0.391\pm0.078$ &  $0.394\pm0.080$ & $0.354\pm0.206$ & $-0.015\pm0.003$ & $-0.015\pm0.003$ & $-0.014\pm0.007$ & $0.101\pm0.065$ & $1.192\pm0.779$ & $-0.15\pm0.08$ \\ 
\hline
\multirow{5}{*}{$\perp$} 
 & -9  & $0.500\pm0.214$ & $0.497\pm0.213$ & -- & $0.505\pm0.211$ & $0.504\pm0.211$ & -- & $0.522\pm5.303$ & $-6.658\pm6.416$ &  --\\ 
\cline{2-11}
 & -7  & $0.510\pm0.218$ & $0.507\pm0.218$ & $0.417\pm0.112$ & $0.516\pm0.215$ & $0.515\pm0.215$ & $0.470\pm0.127$ & $0.447\pm4.794$  &$-7.824\pm6.505$ & $-0.2\pm5.6$  \\ 
\cline{2-11}
 & -5  & $0.521\pm0.222$ & $0.517\pm0.222$ & $0.426\pm0.115$ & $0.527\pm0.219$ & $0.525\pm0.219$ & $0.456\pm0.123$ & $0.374\pm4.125$  &$-9.261\pm6.742$ & $-0.2\pm5.1$  \\ 
\cline{2-11}
 & -3  & $0.530\pm0.227$ & $0.527\pm0.227$ & $0.433\pm0.121$ & $0.537\pm0.223$ & $0.535\pm0.223$ & $0.441\pm0.119$ &  $0.318\pm3.238$ & $-11.060\pm7.239$ & $-0.1\pm4.3$  \\ 
\cline{2-11}
 & -1  & $0.539\pm0.231$ & $0.536\pm0.231$ & $0.440\pm0.123$ & $0.547\pm0.227$ & $0.545\pm0.226$ & $0.423\pm0.114$ & $0.302\pm2.049$  & $-13.364\pm8.171$ & $-0.2\pm2.9$  \\ 
\hline
\multirow{5}{*}{$\parallel$} 
 & -9  & $0.529\pm0.219$ & $0.529\pm0.219$ & -- & $0.496\pm0.209$ & $0.494\pm0.209$ & -- & $2.123\pm5.381$ & $-6.440\pm6.231$ &  --\\ 
\cline{2-11}
 & -7  & $0.544\pm0.224$ & $0.543\pm0.223$ & $0.458\pm0.119$ & $0.508\pm0.214$ & $0.506\pm0.214$ & $0.294\pm0.080$ &$2.017\pm4.876$  &$-7.621\pm6.316$  & $-2.4\pm5.8$  \\ 
\cline{2-11}
 & -5  & $0.559\pm0.229$ & $0.559\pm0.229$ & $0.472\pm0.123$ & $0.520\pm0.218$ & $0.518\pm0.218$ & $0.314\pm0.085$ & $1.855\pm4.208$  &$-9.087\pm6.563$ & $-2.3\pm5.3$  \\ 
\cline{2-11}
 & -3  & $0.575\pm0.233$ & $0.575\pm0.233$ & $0.488\pm0.127$ & $0.532\pm0.222$ & $0.530\pm0.222$ & $0.335\pm0.087$ & $1.616\pm3.319$  & $-10.933\pm7.097$  & $-2.1\pm4.4$  \\ 
\cline{2-11}
 & -1  & $0.592\pm0.238$ & $0.591\pm0.237$ & $0.503\pm0.131$ & $0.545\pm0.227$ & $0.543\pm0.226$ & $0.357\pm0.093$ & $1.259\pm2.122$ & $-13.313\pm8.104$ & $-1.8\pm3.1$  \\ 
\hline\hline
\end{tabular}
\end{adjustbox}
\caption{Local vector ($\mathcal{F_\lambda}$) and tensor ($\mathcal{F}_{\lambda,T}$) form factor, and non-local ($\Vff{\lambda}$) form factor estimates obtained from LCSRs at different values of $q^2$ using inputs from Table~\eqref{tab:inputs}. We compare our predictions with Ref.\cite{Gubernari:2020eft} in the last column of each form factors which includes up to twist-4 contributions.}
\label{tab:FF_values}
\end{table}

The non-local form factors $\Vff{\lambda}$ are then calculated at several $q^2$ values in the spacelike region, namely $q^2=\{-9,-7,-5,-3,-1\}\,\gev^2$, incorporating the twist-5 and twist-6 contributions of the three-particle $B_s$-meson LCDAs and are compared with the case where only up to twist-4 effects are included. The corresponding estimates for these are shown in Table~\eqref{tab:FF_values} where we notice that the total contribution including twist-5 and twist-6 is enhanced by an order of magnitude compared to the case, which includes effects only up to twist-4. Furthermore, we compare our predictions with Ref.\cite{Gubernari:2020eft} in the last column of each form factors. The slight difference between our twist-4 results and those of Ref.~\cite{Gubernari:2020eft} arises from the use of different input parameters. As mentioned in Secs.~\ref{sec:nonlocalFF} and \ref{sec:localFF}, we have reproduced the analytical expressions of the form factors exactly up to twist-4 contributions and results for the coefficients are provided in the ancillary file. 

In order to understand the contributions arising from different twists we provide a breakup at a benchmark value of $q^2=-1\,\gev^2$ as
\begin{equation}
\begin{aligned}
   10^7 \times \Vff{\perp}(-1)&=1.536\vert_3 - 1.235 \vert_4 - 14.334 \vert_5 + 0.668 \vert_6\,,\\
    10^7 \times \Vff{\parallel}(-1)&=1.524 \vert_3 - 0.265 \vert_4 - 14.215 \vert_5 - 0.358 \vert_6\,,\\
     10^7 \times \Vff{0}(-1)&=-0.048 \vert_3 + 0.148 \vert_4 + 0.842 \vert_5 + 0.249 \vert_6\,,
\end{aligned}
\end{equation}
where the subscript denotes different twists ranging from twist-3 to twist-6. It should be noted that we have obtained the same relative signs between different twists at several $q^2$-values for each non-local form factor $\Vff{\lambda}$, resulting in an additive effect on the overall contribution. There is a cancellation between twist-3 and twist-4 contributions for all $\Vff{\lambda}$, as reported in Ref.~\cite{Gubernari:2020eft}, which is significant in the perpendicular case. The new higher-twist corrections are primarily driven by the twist-5 contributions. Notably, except for the perpendicular case, these corrections have the same sign.

\begin{table}[t]
\begin{adjustbox}{max width=\textwidth}
    \centering
    \renewcommand*{\arraystretch}{1.5}
    \begin{tabular}{||c|c||c||c|c|c||c|c|c||c||c||}
    \hline\hline
    \multirow{2}{5em}{\centering Order} & 
  \multirow{2}{8em}{\centering Traditional DAs} & 
    \multicolumn{8}{c||}{LCDA contributions $(10^{-8})$}&\multirow{2}{2em}{Total}\\ \cline{3-10}
    & &  Twist-3& \multicolumn{3}{c||}{Twist-4}& \multicolumn{3}{c||}{Twist-5}& twist-6&   \\ \hline \hline
 \multirow{2}{3em}{$\mathcal{O}\left(\frac{x}{x\cdot v}\right)^0$} & $\psi _A-\psi _V $ & $\cellcolor{red!25} 54.86_{\phi _3} $ &  &  &  &  &  &  & & 54.86\\  & $
 \psi _V $ & $\cellcolor{red!25} -36.98_{\phi _3} $ & $\cellcolor{green!25} -454.11_{\phi _4} $ &  &  &  &  &  & &-491.09\\ \hline\hline
 \multicolumn{2}{||c||}{Total at $\mathcal{O}\left(\frac{x}{x\cdot v}\right)^0 $} &17.88&-454.11& & & & & & & -436.23\\ \hline
  \multicolumn{2}{||c||}{Total from twists at  $\mathcal{O}\left(\frac{x}{x\cdot v}\right)^0$ } & 17.88& \multicolumn{3}{c||}{-454.11}& \multicolumn{3}{c||}{}& & \\ \hline\hline
 \multirow{4}{3em}{$\mathcal{O}\left(\frac{x}{x\cdot v}\right)^1$}
  & $\overline{X}_A $ & $\cellcolor{red!25} -4.35_{\overline{\phi }_3} $&  $\cellcolor{green!25} 54.29_{\overline{\phi }_4} $ & $\cellcolor{blue!25} -7.61_{\overline{\psi }_4} $ &  & &  &  & & 42.33\\
 & $
 \overline{Y}_A+\overline{W} $ & $\cellcolor{red!25} -32.62_{\overline{\phi }_3} $&  & & $\cellcolor{blue!25} 77.40_{\overline{\tilde{\psi }}_4} $& & $ \cellcolor{green!25} -139.47_{\overline{\tilde{\phi }}_5} $&  $\cellcolor{blue!25} 323.30_{\overline{\tilde{\psi }}_5} $ &  & 228.61 \\  & $
 \overline{\tilde{X}}_A $ & $\cellcolor{red!25} -0.33_{\overline{\phi }_3} $ & $ \cellcolor{green!25}-4.68_{\overline{\phi }_4} $& & $\cellcolor{blue!25} 1.61_{\overline{\tilde{\psi }}_4} $ &    &  &  & &-3.40 \\  & $
 \overline{\tilde{Y}}_A $ & $\cellcolor{red!25} 33.94_{\overline{\phi }_3} $ & $ \cellcolor{green!25} 416.49_{\overline{\phi }_4} $ & & $\cellcolor{blue!25} -80.43_{\overline{\tilde{\psi }}_4} $&& & $\cellcolor{blue!25} -335.84_{\overline{\tilde{\psi }}_5} $ & &  34.16\\ \hline\hline
 \multicolumn{2}{||c||}{Total at $\mathcal{O}\left(\frac{x}{x\cdot v}\right)^1 $} &-3.36& 466.10& -7.61&-1.42 & & -139.47&  -12.54&  & 301.70\\ \hline
\multicolumn{2}{||c||}{Total from twists at $\mathcal{O}\left(\frac{x}{x\cdot v}\right)^1$ } & -3.36& \multicolumn{3}{c||}{457.07}& \multicolumn{3}{c||}{-152.01}& & \\[1ex] \hline\hline
 \multirow{2}{3em}{$\mathcal{O}\left(\frac{x}{x\cdot v}\right)^2$} & $
 \overline{\overline{W}} $&  & $\cellcolor{green!25} -31.86_{\overline{\overline{\phi }}_4} $ & $\cellcolor{blue!25} 2.34_{\overline{\overline{\psi }}_4} $ & $\cellcolor{blue!25} 4.74_{\overline{\overline{\tilde{\psi }}}_4} $ & $\cellcolor{blue!25} 11.43_{\overline{\overline{\psi }}_5} $ & $ \cellcolor{green!25} -7.72_{\overline{\overline{\tilde{\phi }}}_5} $ & $\cellcolor{blue!25} 23.12_{\overline{\overline{\tilde{\psi }}}_5} $ &  &2.05 \\  & $
 \overline{\overline{Z}} $ & $\cellcolor{red!25} 0.84_{\overline{\overline{\phi }}_3} $ & $\cellcolor{green!25} 13.45_{\overline{\overline{\phi }}_4} $ & & $\cellcolor{blue!25} -4.00_{\overline{\overline{\tilde{\psi }}}_4} $& & $ \cellcolor{green!25} 3.62_{\overline{\overline{\tilde{\phi }}}_5} $ & $\cellcolor{blue!25} -21.76_{\overline{\overline{\tilde{\psi }}}_5} $ & $\cellcolor{red!25} 6.68_{\overline{\overline{\phi }}_6} $ & -1.17 \\ \hline\hline
 \multicolumn{2}{||c||}{Total at $\mathcal{O}\left(\frac{x}{x\cdot v}\right)^2 $}& 0.84& -18.41& 2.34& 0.74& 11.43& -4.1& 1.36& 6.68& 0.88\\ 
 \hline
\multicolumn{2}{||c||}{Total from twist at $\mathcal{O}\left(\frac{x}{x\cdot v}\right)^2$ } & 0.84& \multicolumn{3}{c||}{-15.33}& \multicolumn{3}{c||}{8.69}&6.68 & \\ \hline \hline
 \multicolumn{2}{||c||}{Total from individual LCDA} & 15.36& -6.42& -5.27& -0.68& 11.43& -143.57& 11.18& 6.68& \\ \hline
 \multicolumn{2}{||c||}{Total from definite twists} & 15.36& \multicolumn{3}{c||}{-12.37}& \multicolumn{3}{c||}{-143.32}& 6.68& -133.65\\ \hline\hline
 \end{tabular}
    \end{adjustbox}
    \caption{Contribution to the non-local form factor $\Vff{\perp}$ at $q^2=-1\,\GeV^2$ arising from various LCDAs with definite twists and distribution amplitudes associated with different Lorentz structures. The LCDAs parametrized in Exponential Model highlighted in green are $\propto |\lambda_E^2+\lambda_H^2|$, in blue are $\propto |\lambda_{E/H}^2|$ and in red are $\propto |\lambda_E^2-\lambda_H^2|$ (see texts for details).}
    \label{tab:FFNL_perp}
\end{table}

A detailed breakup of the contributions to non-local form factors $\Vff{\perp}$ at $q^2=-1\,\gev^2$, separating the LCDAs with definite twist and various traditional distributions amplitudes associated with different Lorentz structures $(\psi_{A,V},\,\ptwiddle{X}_{A})\,\ptwiddle{Y}_{A},\,W,\,Z)$, is provided in Table~\eqref{tab:FFNL_perp} which are arranged in the orders of $1/(v\cdot x)$. For a fast-moving meson $v\cdot x \rightarrow \infty$ and hence denoting the order as leading, next-to leading and next-to-next-to leading order effects. Due to the different signs of the coefficients $C^{(\mathcal{L},\psi_\text{3p})}_{n,r}$ in the LCSR of the non-local form factors (in Eq.~\eqref{eq:CoeffFuncs3pt}), we find a significant amount of the contributions arising from the LCDA of a definite twist are canceled by contributions from the different traditional distribution amplitudes except for the $\phi_4$ and $\tilde{\phi}_5$. There is also a cancellation between contributions arising at different orders of $1/(v\cdot x)$ for $\phi_4$ case.  Also note that these contributions depend on the explicit expressions of the LCDAs for which we used the Exponential Model (see Eq.~\eqref{eq:3_LCDA_Expo}) where $\phi_4$ and $\tilde\phi_5$ are proportional to $|\lambda_E^2+\lambda_H^2|$ (highlighted in green), $\ptwiddle{\psi}_4$ and $\ptwiddle{\psi}_5$ are proportional to $|\lambda_{E/H}^2|$ (highlighted in blue), whereas $\phi_3$ and $\phi_6$ are proportional to $|\lambda_E^2-\lambda_H^2|$ (highlighted in red). As the non-perturbative parameters $\lambda_E^2$ and $\lambda_H^2$  are of the same order, despite of being the leading twist, the contributions from $\phi_3$ see an overall suppression compared to the rest. A similar discussion also follows for the form factors $\Vff{\parallel,0}$ and the corresponding details of various contributions are provided in Appendix~\eqref{app:LCDA_contributions}.

As discussed previously in Sec.~\eqref{sec:localFF}, we also calculate the local form factors relevant for $B_s\rightarrow\phi$ transition incorporating the three-particle twist-5 and twist-6 $B_s$-meson LCDAs. For local form factors, these corrections are less than one percent, due to the dominance of two-particle contributions. The values for the vector and tensor form factors calculated at spacelike region namely at $q^2=\{-9,-7,\dots,-1\}$ $\text{GeV}^2$ are presented in the first two columns of Table~\eqref{tab:FF_values}. 

\subsubsection{Sensitivity to \texorpdfstring{$\lambda_E^2$ and $\lambda_H^2$}{}}

The non-local form factor estimates presented in the previous section are based on the Exponential Model of the $B_s$-meson LCDAs which depends heavily on the following three intrinsic parameters: $\lambda_{B_s}$, $\lambda_E^2$ and $\lambda_H^2$. There have been several attempts to estimate the inverse moment of leading twist LCDA parameter $\lambda_{B_s}$, both directly from QCD sum rule calculations—including systematic inclusion of the strange-quark mass effect~\cite{Khodjamirian:2020hob},—and indirectly by extracting it from Lattice QCD data on form factors~\cite{Mandal:2024pwz}. These two approaches give estimates of $\lambda_{B_s}$ consistent with each other within the uncertainty ranges. However, for the three body quark-gluon parameters $\lambda_{E,H}^2$, the available estimates differ from each other significantly, specifically by an order of magnitude for $\lambda_E^2$.  Due to this reason, we perform a sensitivity study of our prediction of non-local form factors on these parameters $\lambda_{E,H}^2$. We write the expressions as a function of the ratio of the mentioned inputs $R\equiv \lambda_E^2/\lambda_H^2$ at a benchmark point $q^2=-1\GeV^2$ as
\begin{equation}
\begin{aligned}
 \Vff{\perp}(-1) &=\lambda _{H}^2 \big[ (50.416 - 50.416 R)_3 +(-8.247 - 24.772 R) _4 +(-178.114 - 121.170 R) _5 \\&\bs+(21.952 - 21.952 R)_6 \big]\times 10^{-7}\,, \\
\Vff{\parallel}(-1) &=\lambda _{H}^2\big[(50.017-50.017 R) _3+(2.331-13.628 R)_4+(-158.448-156.942 R)_5  \\&\bs+(-11.707+11.707 R)_6  \big]\times 10^{-7}\,,\\
 \Vff{0}(-1)&=\lambda _{H}^2 \big[(-1.566 + 1.566 R)_3 +(2.158 + 0.619 R)_4 +(14.654 - 1.394 R)_5\\&\bs+ (8.179 - 8.179 R) _6\big] \times 10^{-7}\label{eq:VNL_num3}\,,
\end{aligned}
\end{equation}
where the subscript denotes different twist contributions. As expected, each term in the breakup is directly proportional to either $\lambda_H^2$ or $\lambda_E^2$ stemming from the $B_s$-meson LCDA parametrization in the Exponential Model. A similar behavior has also been observed in the case of the Local Duality Model for the LCDAs~\cite{Braun:2017liq} however we refrain from presenting those here.  

\begin{figure}
    \centering
\includegraphics[width=0.47\linewidth]{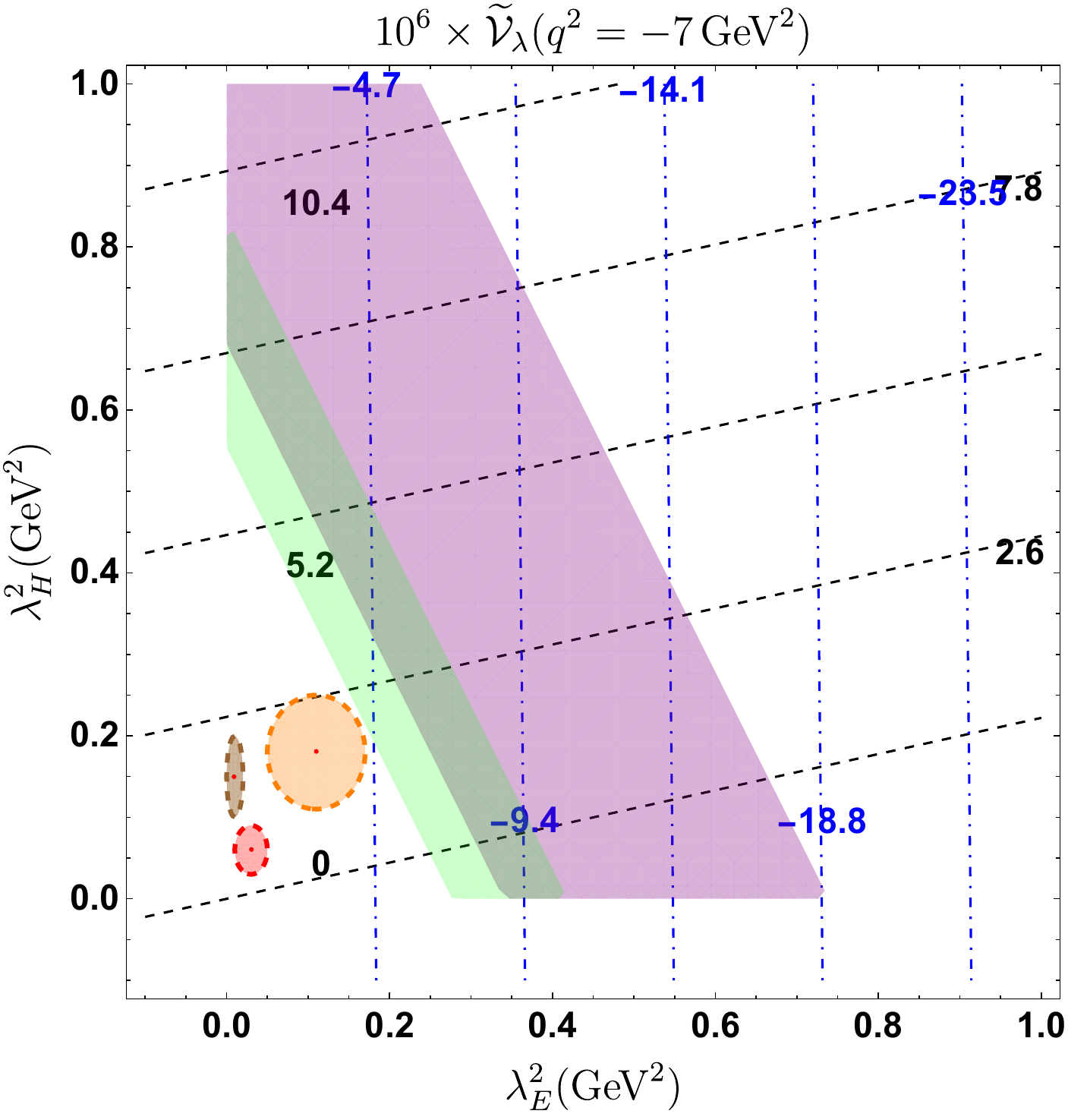}
\includegraphics[width=0.47\linewidth]{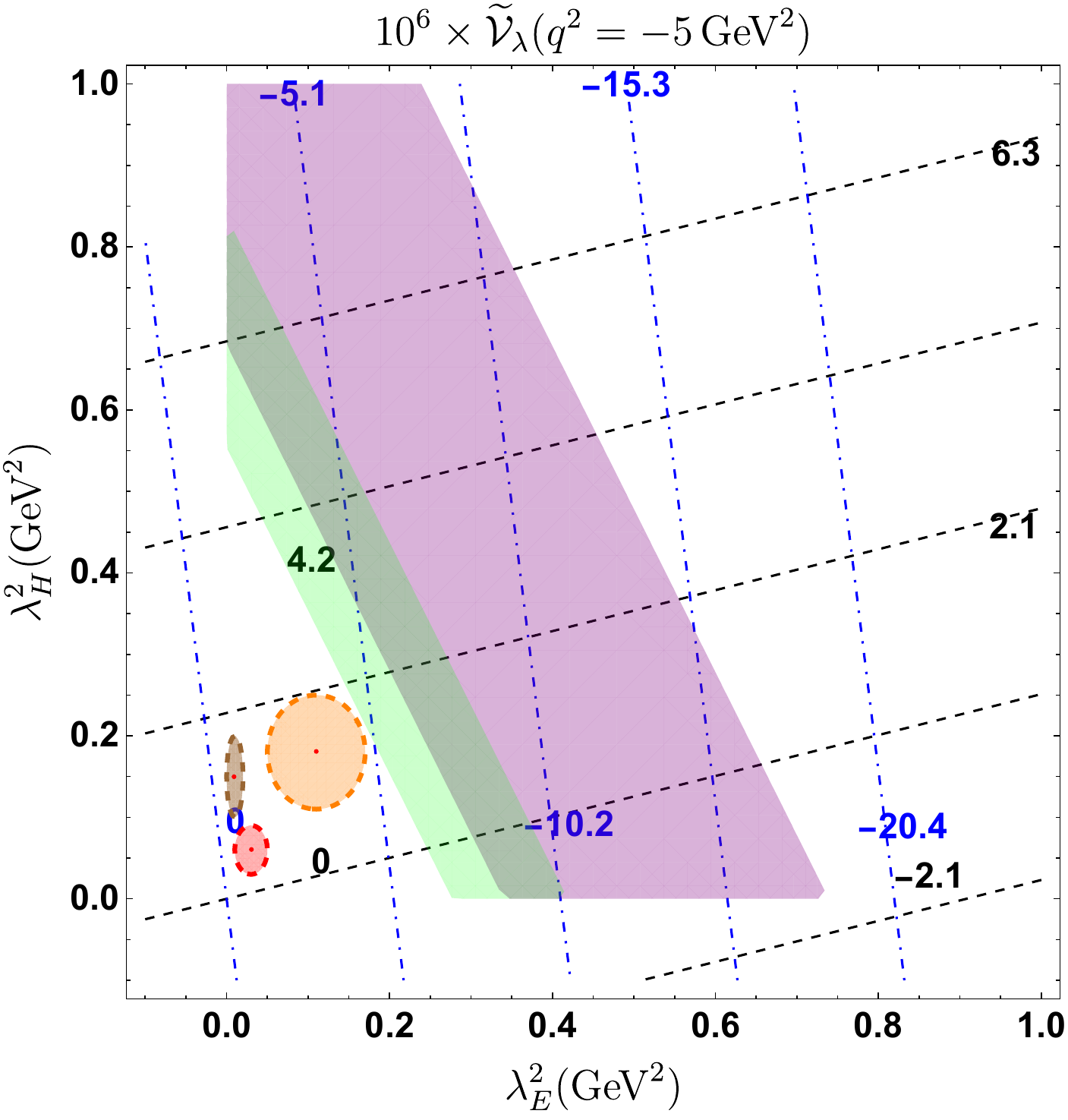}
     \newline
\includegraphics[width=0.47\linewidth]{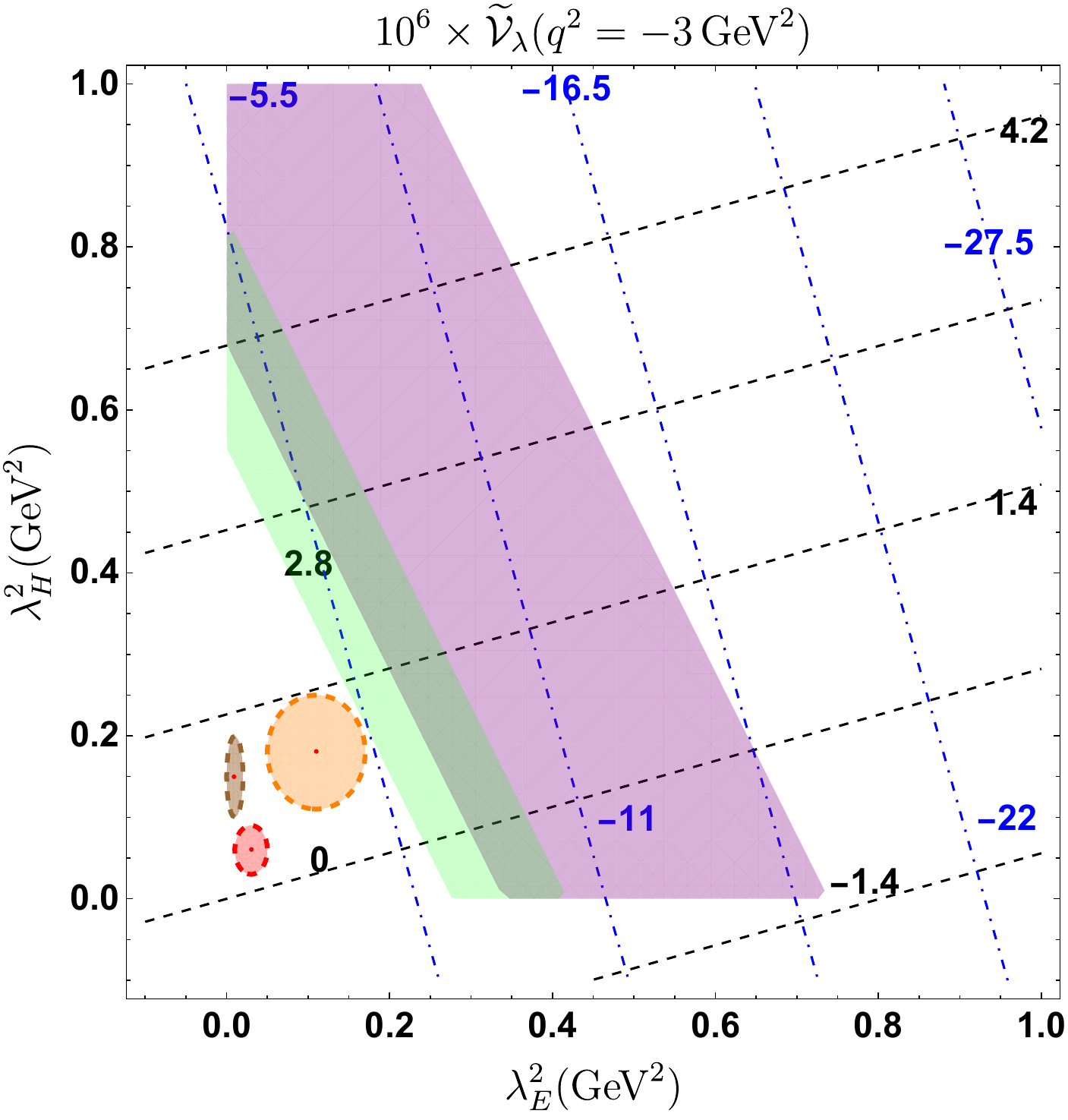}
\includegraphics[width=0.47\linewidth]{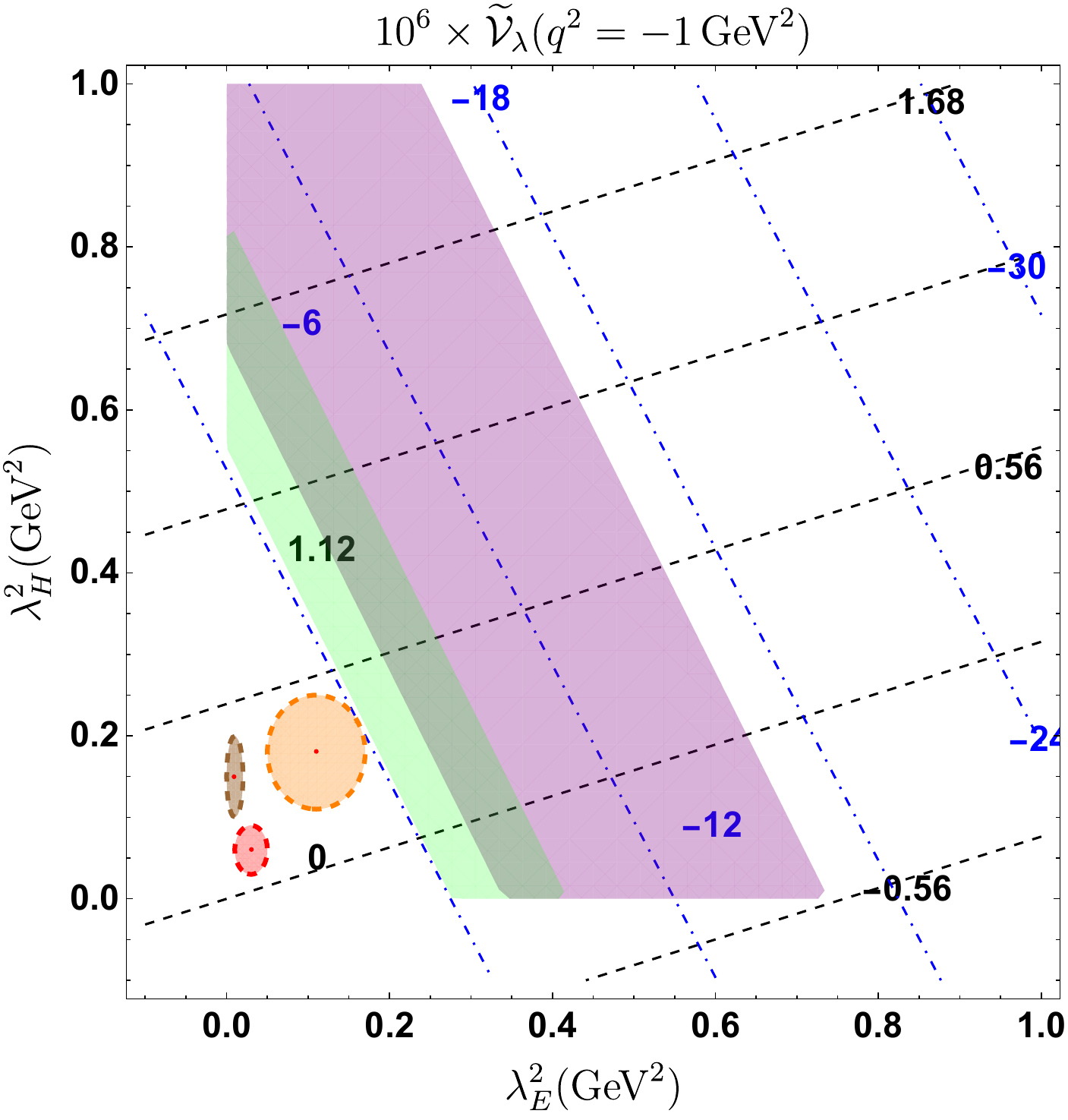} 
    \caption{Contour plots illustrates the dependence of the non-local form factors $\Vff{\lambda}$ on $\lambda_E^2$ and $\lambda_H^2$ across different regions for $q^2=\{-7,-5,-3,-1\}$ $\text{GeV}^2$. The black dashed lines represent $\Vff{0}$, while the blue dashed lines correspond to $\Vff{\perp}$ or $\Vff{\parallel}$, as their numerical values are very similar. The numerical values along these lines indicate the form factor values (in the units of \texorpdfstring{$10^{-6}$}{}). The small red ellipse represents Region I, the larger orange ellipse corresponds to Region II, and the brown ellipse denotes Region III, as described in the text. The shaded pink and green regions correspond to Region IV and Region V, respectively.}
    \label{fig:lambda_dependence}
\end{figure}

We then study the variations in the estimates of the non-local form factors $\Vff{\lambda}$ in Fig.~\eqref{fig:lambda_dependence} in the region $0<\lambda_{E,H}^2<1$ at four different $q^2$ values. The black dashed lines correspond to $\Vff{0}$ values and the blue dashed lines correspond to $\Vff{\perp}$ and $\Vff{\parallel}$ estimates which are numerically very close. As mentioned earlier, several different determinations and theoretical constraints are available in the literature for $\lambda_{E,H}^2$, which are labeled as different regions divided into five categories as follows. 

\begin{itemize}
    \item Region I shown in the tiny red ellipse is the $1\sigma$ region for the QCD sum rule estimates $\lambda_{E}^2=0.03\pm0.02\GeV^2$ and $\lambda_{H}^2=0.06\pm0.03\GeV^2$ provided in Ref.~\cite{Nishikawa:2011qk} and also used in the numerical analysis of this article. The authors included several $\mathcal{O}(\alpha_s)$ corrections to non-perturbative entries in the sum rule.
    \item Region II, the $1\sigma$ region obtained for slightly higher estimates provided in Ref.~\cite{Grozin:1996pq} namely $\lambda_{E}^2=0.11\pm0.06\GeV^2$ and $\lambda_{H}^2=0.18\pm0.07\GeV^2$, is depicted by relatively large orange circle. These are one of the first predictions based on the sum rules calculated within the HQET approach, where the authors restricted the local vacuum condensates up to mass dimension five in the OPE calculation.
     \item Region III depicted by brown ellipse is the $1\sigma$ region obtained for the estimates given in Ref.~\cite{Rahimi:2020zzo} that read: $\lambda_{E}^2=0.01\pm0.01\GeV^2$ and $\lambda_{H}^2=0.15\pm0.05\GeV^2$. Here the authors used an alternative sum rule starting with a correlation function with two three-particle quark-gluon-antiquark currents.
\item Region IV denoted by pink shaded area is allowed by the Grozin-Neubert~\cite{Grozin:1996pq} relation for the Exponential Model, i.e., $2\bar{\Lambda}^2=2\lambda_E^2+\lambda_H^2$~\cite{Lu:2018cfc} where $\bar{\Lambda}$ is approximated with the binding energy $\bar{\Lambda}=m_{B_s}-M_b$ in the vanishing strange quark mass limit and $M_b$ is the pole mass definition of the bottom quark. 
\item Region V shown in green shaded area is obtained using the same constraint as of Region IV. However here,  $\bar{\Lambda}$ is related to the inverse moment of the $B_s$-meson distribution amplitude with the use of equations of motion for the distribution amplitudes. This  results in  $\bar{\Lambda}=3/2\,\lambda_{B_s}$ and the Grozin-Neubert relation in this approximation is $9/2\lambda_{B}^2=2\lambda_E^2+\lambda_H^2$.
\end{itemize}

We note a significant variation in the non-local form factor $\Vff{\lambda}$ estimates relative to Region I.
This includes an enhancement of 
 $\sim$ $200\%$ for $\mathcal{\widetilde{V}}_{0}$ in Region II and  Region III, and $\sim50\%$ in the case of $\mathcal{\widetilde{V}}_{\perp,\parallel}$. Along the parameter space in Regions IV and V, such estimates can even vary by several orders of magnitude which are shown by the contour lines. It is clear from Fig.~\eqref{fig:lambda_dependence} that much precise estimates of the inputs $\lambda_{E,H}^2$ will be crucial for the non-local form factor determination. 

\subsection{Form factors in timelike region}\label{sec:fitdetails}
The estimation of form factors in the timelike region is necessary in order to make predictions for the observables that are measured at experiments within this physical region. In this section, we discuss the methods used to extrapolate the LCSR predictions for both the local and the non-local form factors into the timelike region. 
\subsubsection{Local form factors}
We now use the standard procedure of analytic continuation with the $z$-expansion in order to extrapolate the local form factors\footnote{Note that the local form factors in the traditional basis are denoted here as $F(q^2)$ and calligraphic font, $\mathcal{F}_\lambda(q^2)$, is used for form factors in the transversity basis.} calculated at the deep spacelike region to the timelike region, which is the kinematically allowed region of the decay. This is achieved by fitting the LCSR predictions to the following ansatz~\cite{Bharucha:2015bzk}:
\begin{equation}
\label{eq:local_zfit}
    F^{(z)}(q^2)\equiv\frac{1}{1-q^2/m^2_{R}}\sum_{i=0}^2\hs a_i^F\hs (z(q^2)-z(0))^i\,,
\end{equation}
where $m_{R}$ denotes the mass of the sub-threshold resonances compatible with the quantum numbers of the respective
form factors and $a^F_i$ are the coefficients of the $z$-expansion. The conformal variable $z(t)$ is given by:
\begin{equation}
    z(t)=\frac{\sqrt{t_+ -t}-\sqrt{t_+-t_0}}{\sqrt{t_+ -t}+\sqrt{t_+-t_0}}\,,
    \label{eq:z_var}
\end{equation}
where $t_\pm=(m_{B_s}\pm m_\phi)^2$ and $t_0=t_+ \left(1-\sqrt{1-t_-/t_+}\right)$ are used. The parameters $t_0$ and $t_+$ are chosen such that $t=t_0$ and $t=t_+$ are mapped to $z=0$ and $\abs{z}= 1$, respectively. The values of the resonance masses relevant for the local form factors used in our study are tabulated in Table~\eqref{tab:m_res}.

  \begin{table}[H] \centering  \begin{tabular}{||c||c||} \hline\hline Form factors& \makecell{Resonance mass\\  ($m_R$ in  MeV)}    \\ \hline\hline   $V$, $T_1$& $5415.40\pm1.40$\\ \hline   $A_{1}$, $A_{2}$, $T_{2}$, $T_{3}$ &$5828.73\pm 0.20$\\ \hline  $A_0$& $5366.93\pm 0.01$\\ \hline\hline  \end{tabular}  \caption{Resonance masses from PDG\cite{ParticleDataGroup:2022pth} required for the analytic continuation of the local form factors in the timelike region.} \label{tab:m_res} \end{table}
The local form factors in the timelike region are obtained by determining the unknown $z$-expansion coefficients $a^F_i$ in Eq.~\eqref{eq:local_zfit} from the LCSR predictions. To achieve this, we minimize the $\chi^2$ for the local form factors defined as:
\begin{align}
    \chi_{F}^2&=\sum_{q^2_i}\left(\frac{F^{(z)}(q_i^2)-F^{\text{LCSR}}(q_i^2)}{\sigma_{\scriptscriptstyle F}(q_i^2)}\right)^2+\chi^2_\text{nuis.}, ~~{\rm where}~~
    \chi^2_\text{nuis.}=\sum_j \left(\frac{m_j-m_j^\text{exp.}}{\sigma_{m_j^\text{exp.}}}\right)^2.
\end{align}
Here, $F^{\text{LCSR}}(q_i^2)$ are the LCSR predictions for the local form factors at $q_i^2\in \{-9,-8,\cdots,-1\}\GeV^2$ with $1\sigma$ uncertainty $\sigma_{\scriptscriptstyle F}(q_i^2)$. The meson masses $m_j\in\{m_{B_s},m_\phi, m_R\}$ are considered as the nuisance parameters and their experimental values with uncertainties are taken from PDG~\cite{ParticleDataGroup:2022pth}. Our best-fit values including $\pm1\sigma$ uncertainties for the coefficients $a_i^F$ are presented in Table~\eqref{tab:z_coef}. The correlation matrix for the $z$-expansion coefficients for all the seven local form factors in the traditional basis is also quoted in Appendix~\eqref{app:num_local_FF}.

\begin{table}[t]
    \centering
    \renewcommand*{\arraystretch}{1.2}
    \begin{tabular}{||c|c|c|c||}\hline\hline
    Form factor ($F$) &$a_0^{F}$& $a_1^{F}$& $a_2^{F}$\\ \hline\hline
 $V$ &$ 0.47\pm 0.20$ &  $ -0.90\pm 0.48$ &  $ -1.80\pm 0.61$ \\ \hline
 $A_1$& $ 0.36\pm 0.14$ &  $ 0.89\pm 0.48$ &  $ -1.08\pm 0.54$ \\ \hline
 $A_2$& $ 0.33\pm 0.17$ &  $ -0.24\pm 0.36$ & $ -3.03\pm 0.34$ \\ \hline
 $T_1$ & $ 0.40\pm 0.17$  & $ -0.82\pm 0.35$ &  $ -1.16\pm 0.48$ \\ \hline
 $T_2$&  $ 0.40\pm 0.17$ & $ 1.44\pm 0.61$  & $ -2.74\pm 0.60$ \\ \hline
 $T_3$ & $ 0.29\pm 0.16$ &  $ 0.02\pm 0.40$ & $ -5.03\pm 0.33$ \\ \hline
 $A_0$& $ 0.42\pm 0.08$ &  $ -1.11\pm 0.23$  & $ 0.19\pm 0.47$ \\ \hline\hline
       \end{tabular}
    \caption{Best fit results for the $z$-expansion coefficients given in Eq.~\eqref{eq:local_zfit} for the local form factors.}
    \label{tab:z_coef}
\end{table}

\begin{table}[h]
    \centering
    \renewcommand{\arraystretch}{1.2}
    \setlength{\tabcolsep}{6pt} 
    \begin{tabular}{||c|c|c|c|c||}
        \hline\hline
        Form factor($F$) & This Work & Ref. \cite{Bharucha:2015bzk} & Ref. \cite{Gubernari:2022hxn} & Ref. \cite{Gubernari:2023puw} \\ \hline 
        \hline
        $V$      & $0.49 \pm 0.21$  & $0.36 \pm 0.04$  & $0.40 \pm 0.05$ & $0.40 \pm 0.05$ \\ \hline
        $A_1$    & $0.36 \pm 0.14$  & $0.28 \pm 0.03$  & $0.31 \pm 0.03$ & $0.31 \pm 0.03$ \\ \hline
        $A_{12}$ & $0.27 \pm 0.05$  & $0.26 \pm 0.03$  & $0.25 \pm 0.03$ & $0.25 \pm 0.02$ \\ \hline
        $T_1$    & $0.42 \pm 0.18$  & $0.30 \pm 0.03$  & $0.36 \pm 0.04$ & $0.35 \pm 0.04$ \\ \hline
        $T_2$    & $0.40 \pm 0.17$  & $0.29 \pm 0.03$  & $0.35 \pm 0.04$ & $0.34 \pm 0.02$ \\ \hline
        $T_{23}$ & $0.70 \pm 0.14$  & $0.69 \pm 0.08$  & $0.70 \pm 0.40$ & $0.65 \pm 0.04$ \\ \hline
        $A_0$    & $0.44 \pm 0.09$  & $0.38 \pm 0.05$  & $0.40 \pm 0.04$ & $0.40 \pm 0.04$ \\ \hline
        \hline
    \end{tabular}
    \caption{Comparison of local form factors at $q^2 = 1 \GeV^2$ with previous results in the literature.}
    \label{tab: LFFs}
    \label{tab:local_form_factors}
\end{table}

We compare our LCSR results for the form factors at $q^2 =1 \GeV^2 $ with previous results in the literature~\cite{Bharucha:2015bzk,Gubernari:2022hxn,Gubernari:2023puw} in Table~\eqref{tab: LFFs}. Our predictions have similar central values, but larger uncertainties largely due to the use of $B_s$-meson LCDAs. Note that in Ref.~\cite{Bharucha:2015bzk}, the local form factors were calculated using light-meson distribution amplitudes which are significantly more precise. In Ref.~\cite{Gubernari:2022hxn,Gubernari:2023puw}, $B_s$-meson LCDAs were employed and the form factors were obtained by simultaneously fitting the $z$-expansion to the LCSR results as well as the Lattice QCD results, resulting in reduced uncertainties.

\subsubsection{Non-local form factors } \label{sec:non-local FF}
The double subtracted dispersion relation ansatz for the non-local transversity amplitude $\mathcal{H}_\lambda(q^2)$ discussed in Sec.~\eqref{sec:disp} requires various experimental and theoretical inputs. On the experimental side, polarization fractions and phases relative to longitudinal polarization are provided in the LHCb studies~\cite{LHCb:2023exl,LHCb:2023sim,LHCb:2016tuh}. The phases for the longitudinal polarization amplitude for different resonances are treated as free parameters in our analysis which represents the relative phase for the resonance channel $B_s \to \phi V$ compared to the non-resonant $B_s \to \phi \bar\ell \ell$ background, and will be determined in our analysis by fitting the dispersion relation to the LCSR predictions. On the theoretical side, updated values of the decay constants, meson masses, and widths are used in the dispersion relation. The decay constant of the $\phi$-meson, $f_\phi$ is taken from Ref.~\cite{Bharucha:2015bzk} where mixing between $\rho$- and $\omega$- mesons with QCD effects are also considered in the determination. The decay constants for the charmonium states are obtained using updated values of their partial decay width to an electron-positron pair from PDG~\cite{ParticleDataGroup:2022pth} and the following expression~\cite{Hatton:2020qhk,Bharucha:2015bzk}.
\begin{equation}
    \Gamma(V\rightarrow e^+ e^-)=\frac{4\pi}{3}\alpha^2_{\text{EM}} Q_c^2 \frac{f_V^2}{m_V}+\mathcal{O}\left(\alpha_\text{EM},\frac{m_V^2}{M_W^2}\right)\,,\label{eq:f_decay}
\end{equation}
where $\alpha_\text{EM}$ at the charmonium masses are obtained from Ref.~\cite{Kuhn:2007vp}. As mentioned earlier, the transversity amplitudes $A_V^\lambda$ for the resonances are obtained by combining the data on polarization fractions and normalization factor related to the decay width of the $B_s\rightarrow\phi V$ mode given in Eq.~\eqref{eq:decay_resonance}.
All these inputs entering in the dispersion relations~\eqref{eq:disppsi} and \eqref{eq:disppsi0} are summarized in Table~\eqref{tab:disp_inputs}.

\begin{table}[h]
 \begin{adjustbox}{max width=\textwidth}
    \centering
    \renewcommand*{\arraystretch}{1.1}
    \begin{tabular}{||c c||c||c||c||c||}
    \hline\hline
   \multirow{2}{*}{Resonance parameter }& &\multirow{2}{*}{Symbol}&\multicolumn{3}{c||}{Resonance (V)}\\ \cline{4-6}
& & &$\phi$ & $J/\psi$ & $\psi(2S)$ \\ \hline\hline
 Masses (in $\MeV$) &\cite{ParticleDataGroup:2022pth} &$m_V$& $1019.461\pm0.016$&$3096.900\pm 0.006$ &
        $3686.097\pm 0.011$\\ \hline \makecell{Total decay widths (in $\KeV$)}&\cite{ParticleDataGroup:2022pth} & $\Gamma^\text{tot.}_{V}$ & $4249\pm13$& 
         $92.6\pm1.7$& $293\pm 9$ \\ \hline
         \makecell{Decay constants (in $\MeV$)}& &$f_V$ &  $233\pm 4$ \cite{Bharucha:2015bzk}&
         $410.4\pm1.7$ & $287\pm 6$   \\ \hline
        \multirow{3}{*}{\makecell{ Amplitudes \\ (in $\GeV^3$)  }}& & $\vert A^0_V\vert$ & $0.016\pm0.001$ & $0.246\pm0.009$ & $0.189\pm0.009$ \\ \cline{3-6}
       & &$\vert A^\parallel_V\vert$ & $0.014\pm0.001$ & $0.166\pm0.006$ & $0.163\pm0.008$ \\ \cline{3-6}
        & &$\vert A^\perp_V\vert$ & $0.014\pm0.001$ & $0.170\pm0.006$ & $0.150\pm0.007$ \\ \hline
          \multirow{2}{*}{\makecell{ Polarization phases}}& & $\varphi^\perp_V$ & $2.77\pm 0.11$~\text{\cite{LHCb:2023exl}} & $2.92\pm 0.08$~\text{\cite{LHCb:2023sim}} & $3.29\pm 0.43$~\text{\cite{LHCb:2016tuh}} \\ \cline{3-6}
 &  &   $\varphi^\parallel_V$ & $2.46\pm 0.03$~\text{\cite{LHCb:2023exl}} & $3.15\pm 0.06$~\text{\cite{LHCb:2023sim}} & $3.67\pm 0.18$~\text{\cite{LHCb:2016tuh}} \\ \hline \hline
    \end{tabular}
    \end{adjustbox}
    \caption{Parameters used in the dispersion relations~\eqref{eq:disppsi} and \eqref{eq:disppsi0} for the non-local transversity amplitudes. The two-body amplitudes \texorpdfstring{$A_{V}^\lambda$}{} are calculated using Eq.~\eqref{eq:decay_resonance} along with the data on polarization fractions of the relevant modes, the decay constant for charmonium states are calculated using Eq.~\eqref{eq:f_decay} and inputs on meson masses, decay width and branching fractions are taken from PDG~\texorpdfstring{\cite{ParticleDataGroup:2022pth}}{}.}
    \label{tab:disp_inputs}
\end{table}
In order to obtain the $\mathcal{H}_{\lambda}$'s in the timelike region, as a next step, we define a $\chi^2$-function similar to the previous subsection as,
\begin{align}
\label{eq:chi2disp}
\chi_{\scriptscriptstyle\mathcal{H}}^2=\sum_{\lambda,\,q^2_i} \left(\frac{\vert\mathcal{H}_{\lambda}^{\text{\tiny{disp.}}} -\mathcal{H}_{\lambda}^{\scriptscriptstyle\text{\tiny{LCSR}}}\vert}{\sigma_{\scriptscriptstyle\mathcal{H}^{\textstyle\text{\tiny LCSR}}_{\lambda}}
}\right)^2_{q^2=q^2_i}+\chi^2_{\scriptscriptstyle\mathcal{H},\text{nuis.}}\,,
\end{align}
and perform the minimization by fitting the LCSR predictions of the three non-local transversity amplitudes, denoted here as $ \mathcal{H}_{\lambda}^{\text{\tiny{LCSR}}}$, in the spacelike regions for  $q_i^2\in \{-9,-8,\cdots,-2\}\GeV^2$ to the double-subtracted dispersion relations given in Eqs.~\eqref{eq:disppsi}-\eqref{eq:disppsi0} represented as $\mathcal{H}_{\lambda}^{\text{\tiny{disp.}}}$. We have taken the subtraction point at $q_0^2 = -1\,\text{GeV}^2$. The fit is optimized for the unknown longitudinal phases $\varphi_V^0$ for the three resonances and the complex parameters $a_\lambda$ and $b_\lambda$ arising in the linear model used to approximate the continuum contributions in the dispersion relations. 

In the $\chi^2$ definition (Eq.~\eqref{eq:chi2disp}), $\sigma_{\scriptscriptstyle\mathcal{H}^{\textstyle\text{\tiny LCSR}}_{\lambda}}
$ is the $1\sigma$ uncertainty in the LCSR estimates of $\mathcal{H}^{\textstyle\text{\tiny LCSR}}_{\lambda}$ and the $\chi^2$ for nuisance parameters, $\mathcal{P}=\{m_{B_s},m_D,s_0,m_V,\Gamma_V,f_V,A^\lambda_V,\varphi_{\parallel,\perp}\}$, is given by
\begin{align}
\chi^2_{\scriptscriptstyle\mathcal{H},\text{nuis.}}=\sum_{x \in \mathcal{P}} \left(\frac{\vert x-x^\text{exp.}\vert}{\sigma_{x^\text{\tiny{exp.}}}}\right)^2\,,\label{eq:H_nuis}
\end{align}
where the central values of the parameters and their $\pm 1\sigma$ uncertainties represented as $x^\text{exp.}\pm\sigma_{x^\text{\tiny{exp.}}}$ are provided in Table~\eqref{tab:inputs} and Table~\eqref{tab:disp_inputs}. The fit results are presented in Table \eqref{tab:fit_comb1}, showing that the relative non-resonant phase $\varphi_V^0$ is significantly smaller for $J/\psi$ and $\psi(2S)$ compared to the $\phi$ resonance. The parameters $a_\lambda$ and $b_\lambda$ exhibit minimal imaginary contributions compared to the corresponding real parts, except in the perpendicular case.

Once these fitting parameters are determined, the dispersion relations~\eqref{eq:disppsi} and \eqref{eq:disppsi0} can be used to evaluate the phenomenological impact of the charm-loop contributions on various angular observables of this decay mode, which will be discussed in the next subsection.

\begin{table}[ht!]
 \begin{adjustbox}{max width=\textwidth}
    \centering
    \renewcommand*{\arraystretch}{1.3}
    \begin{tabular}{||c c ||c c||c c||}
    \hline\hline
\multicolumn{6}{||c||}{ Best fit value of parameters}\\ \hline \hline
\multicolumn{2}{||c||}{$\varphi_\phi^0=-0.301$}&\multicolumn{2}{c||}{ $\varphi_{J/\psi}^0=0.034$}& \multicolumn{2}{c||}{$\varphi_{\psi(2S)}^0=0.018$} \\ \hline
  $\mathrm{Re}(a_{\perp})=$&$(-0.754\pm0.579 )\times 10^{-6}$& $\mathrm{Re}(a_{\parallel})=$&$(-0.944\pm0.637) \times 10^{-6}$& $\mathrm{Re}(a_{0})=$&$(-1.040\pm0.371)\times 10^{-6}$ \\ \hline 
   $\mathrm{Im}(a_{\perp})=$&$ (0.120\pm0.003)\times 10^{-6}$& $\mathrm{Im}(a_{\parallel})= $&$(0.050\pm0.004) \times 10^{-6}$& $\mathrm{Im}(a_{0})=$&$(- 0.067\pm0.003)\times 10^{-6}$ \\ \hline 
   $\mathrm{Re}(b_{\perp})=$&$(-0.098\pm0.328)\times 10^{-6}$& 
  $\mathrm{Re}(b_{\parallel})=$&$(-0.205\pm0.348)\times 10^{-6}$ &
  $\mathrm{Re}(b_{0})=$&$(-0.377\pm0.263 )\times 10^{-6}$\\ \hline
   $\mathrm{Im}(b_{\perp})=$&$(0.077\pm0.003)\times 10^{-6}$& 
  $\mathrm{Im}(b_{\parallel})=$&$(0.037\pm0.003 )\times 10^{-6}$ &
  $\mathrm{Im}(b_{0})=$&$(-0.040\pm0.002)\times 10^{-6}$\\ \hline
    \end{tabular}
     \end{adjustbox}
    \caption{Fit results for the complex parameters used to model continuum contributions in Eq.~\eqref{eq:model_larges} and the non-resonant phases which are obtained by varying them within the range $-\pi<\varphi_{\phi }^0,\varphi_{J/\psi }^0,\varphi_{\psi(2S)}^0<\pi$ in the dispersion relations~\eqref{eq:disppsi} and \eqref{eq:disppsi0}. }
    \label{tab:fit_comb1}   
\end{table}

\subsection{Contributions to \texorpdfstring{$C_9$}{} and impact on the angular observables } 
\label{sec:angobs}

\begin{figure}[t]
    \centering
    \includegraphics[width=0.45\linewidth]{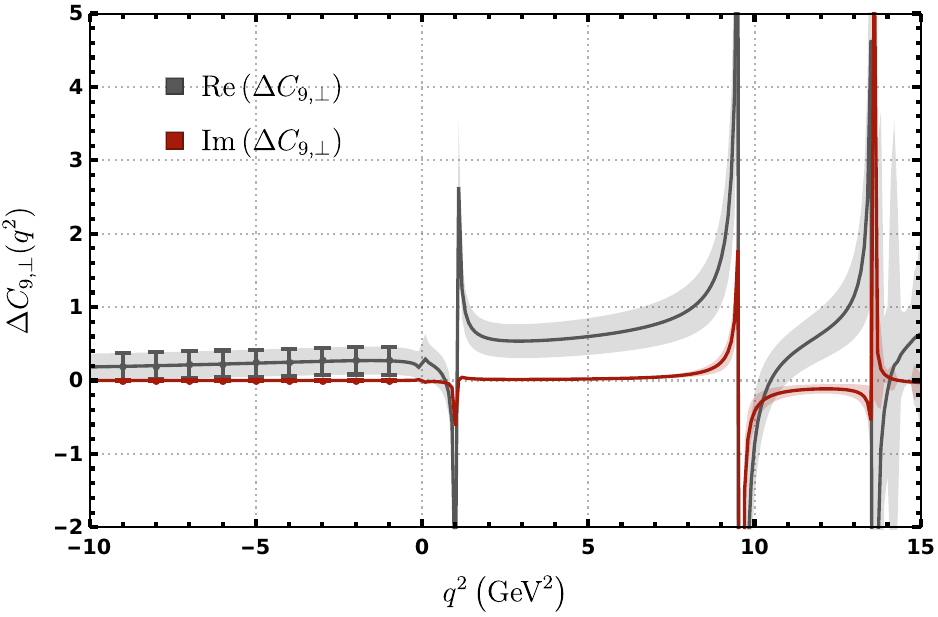}
    \includegraphics[width=0.45\linewidth]{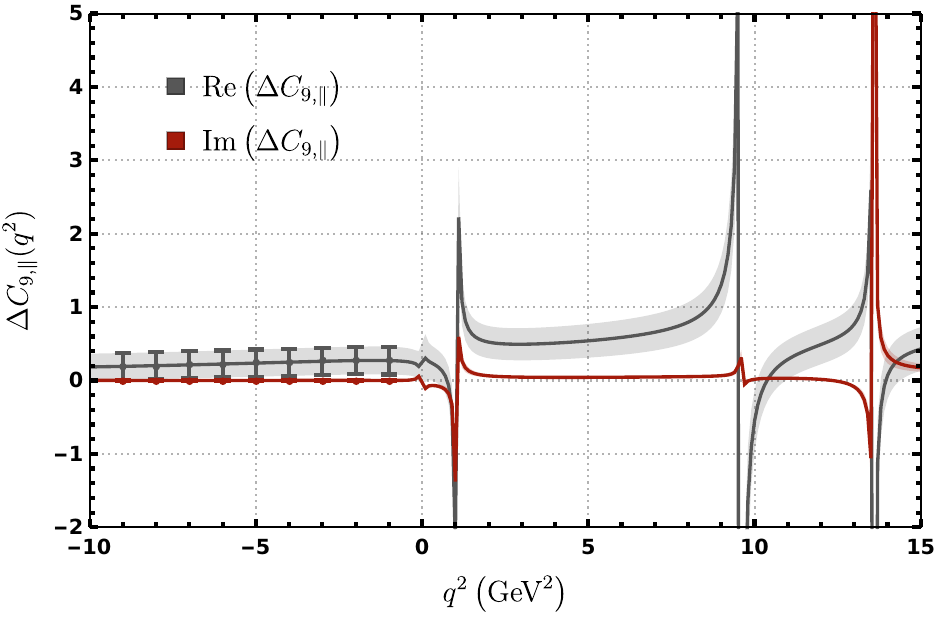}
        \includegraphics[width=0.45\linewidth]{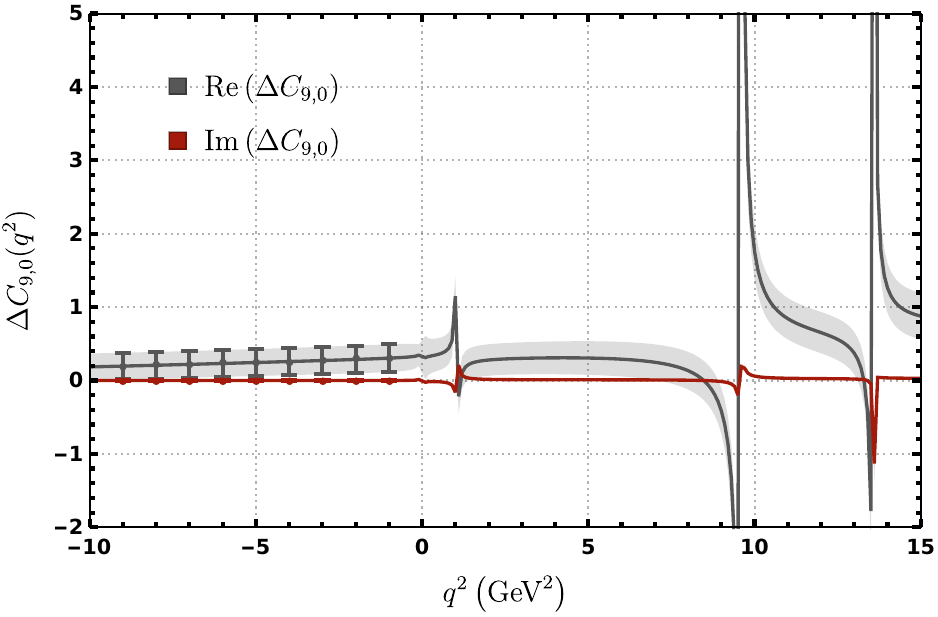}
    \caption{The real and imaginary parts of $\Delta C_{9,\lambda}$ are obtained from the analytic continuation of $\mathcal{H}_\lambda(q^2)$ and $\mathcal{F}_\lambda$ using Eq.~\eqref{eq:c9_defs}, along with the fitted values from Table~\eqref{tab:fit_comb1} in the dispersion relations~\eqref{eq:disppsi}-\eqref{eq:disppsi0}. Local form factors are included from Eq.~\eqref{eq:local_zfit}, incorporating  inputs from Table~\eqref{tab:z_coef}. The shaded band represents $1\sigma$ uncertainties at different $q^2$ values, while the error bars at $q^2<0$ correspond to the LCSR predictions with $1\sigma$ uncertainties. }
    \label{fig:c9}
\end{figure}

The effect of the charm-loop contributions given in Eq.~\eqref{eq:NLFF_full} can be absorbed in the redefinition of the Wilson coefficient $C_{9}^{\text{eff.}}$ for the $B_s \to \phi \bar{\ell} \ell$ transition.
The modified Wilson coefficient for $B_s \to \phi \bar{\ell} \ell$ then can be written as 
\begin{align}
\label{eq:C_9eff1}
C_{9,\lambda}^{\text{eff.}}(q^2)&\equiv C^\text{eff.}_9+\Delta C_9+\Delta C_{9,\lambda}^{\text{NF}}(q^2)=C^\text{eff.}_9+\Delta C_{9,\lambda}(q^2)\,,
\end{align}
where 
\begin{align}
\Delta C_{9,\lambda}(q^2)&=\Delta C_9- \frac{ 32\pi^2m_{B_s}^2}{q^2} \,2 \,Q_c\, \left(C_2 - \frac{C_1}{2N_c} \right)
    \frac{\Vff{\lambda}}{\mathcal{F}_\lambda}\equiv -\frac{32 \pi^2 m_{B_s}^2}{q^2}\frac{\mathcal{H}_{\lambda}}{\mathcal{F}_{\lambda}}\,,
 \label{eq:c9_defs}
\end{align}
includes both the factorizable $\Delta C_9$ and non-factorizable contributions proportional to the non-local form factors $\Vff{\lambda}$. Here, in the last line, we used the notation introduced in Eq.~\eqref{eq:NLFF_full}.
Using the analytic continuation in the timelike region of the non-local transversity amplitude $\mathcal{H}_{\lambda}$ as well as the local form factors $\mathcal{F}_{\lambda}$ obtained in the preceding sections, we can predict such modification in the Wilson coefficient $C_9$ as a function of $q^2$ which is illustrated in Fig.~\eqref{fig:c9}. The real and imaginary parts of $\Delta C_{9,\lambda}$ are shown in gray and red bands, respectively, for all three polarizations. We have also overlaid the predictions of $\Delta C_{9,\lambda}$ from the LCSR calculation in the negative $q^2$ region. The region around $q^2$ $\simeq$ $1\,\gev^2$ shows the presence of the light vector resonance $\phi$. In the range $[2-8]\, \text{GeV}^2$, $\Delta C_{9,\lambda}$ remains well-behaved but grows and exhibits discontinuities as $q^2$ approaches the charmonium resonances $J/\psi$ and $\psi(2S)$. Notably, in all three polarization cases, $\Delta C_{9,\lambda}$ remains positive $(\Delta C_{9,\lambda} >0)$ across most of the $q^2$ region, except the vicinity of the resonances.

\begin{figure}[ht!]
    \centering
    \includegraphics[width=0.46\linewidth]{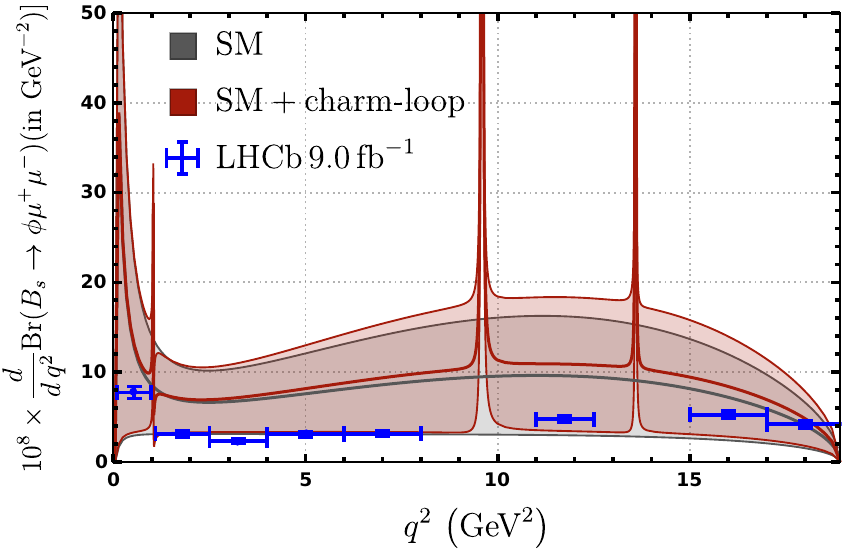}
\includegraphics[width=0.45\linewidth]{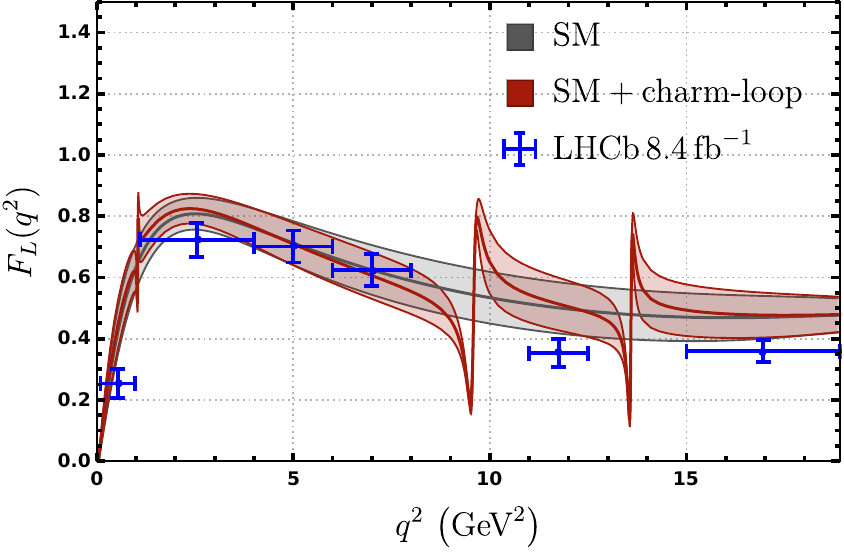}
\includegraphics[width=0.45\linewidth]{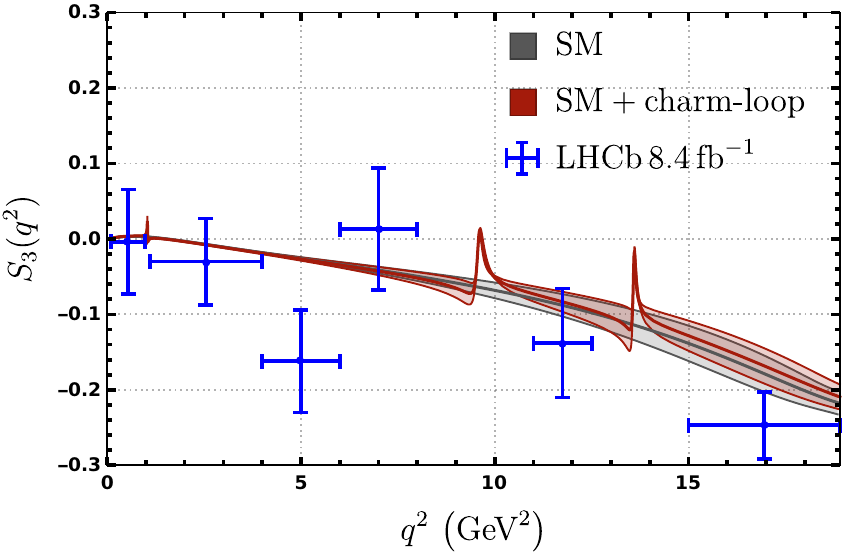}
\includegraphics[width=0.45\linewidth]{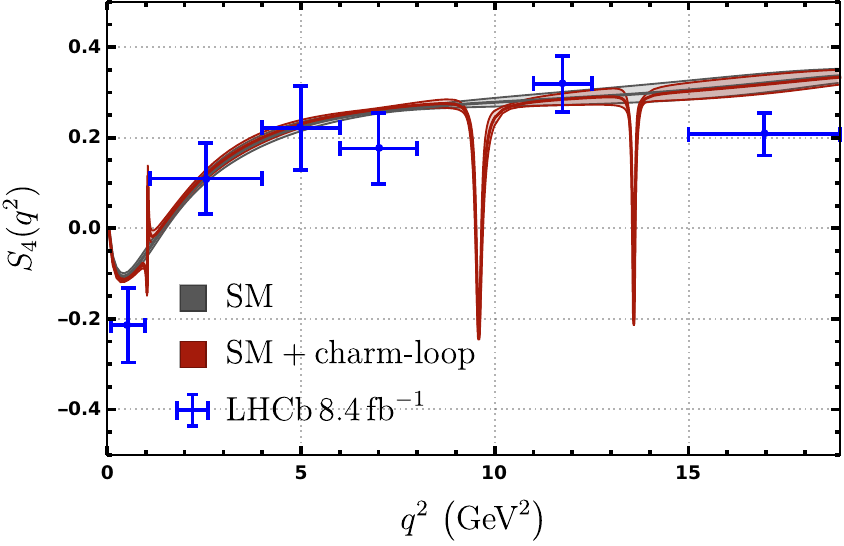}
\includegraphics[width=0.45\linewidth]{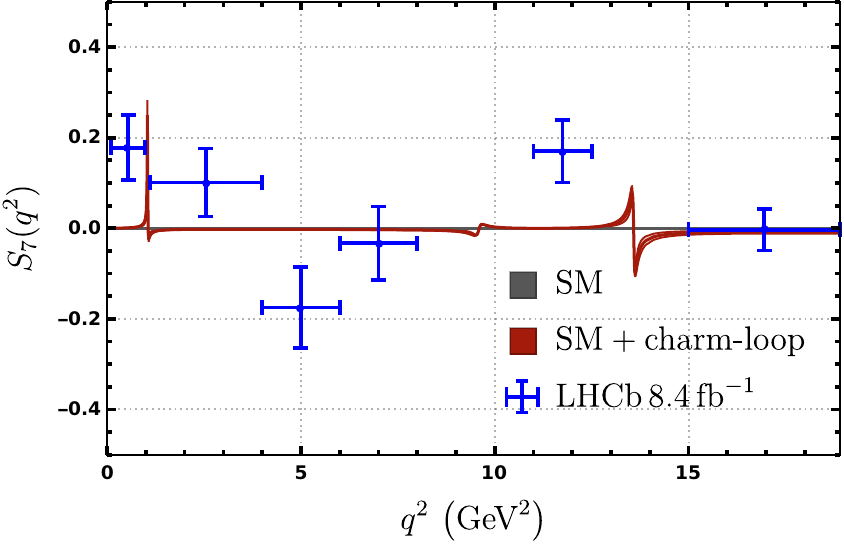}
    \caption{Predictions of the differential branching fraction, longitudinal polarization fraction and the CP-averaged angular observables overlaid with the experimental data~\cite{LHCb:2021zwz,LHCb:2021xxq}. The `SM' predictions, denoted by gray bands, include the factorizable ($\Delta C_9$) corrections to $C_9^{\rm eff.}$ (see Eq.~\eqref{eq:C_9eff1}) whereas the `SM+charm-loop' corrections, depicted by red bands, are obtained by further including the non-factorizable charm-loop corrections ($\Delta C_{9,\lambda}^{\rm NF}$) to the `SM' predictions. The bands represent the $1\,\sigma$ uncertainty regions. The experimental data is shown with blue error bars for several $q^2$-bins. }
    \label{fig:observables}
\end{figure}

Now, we investigate the impact of the charm-loop contributions obtained in this work on the observables of this decay mode. The decay $B_s \to \phi(\to K^+ K^-) \mu^+ \mu^-$ and its $CP$-conjugate $\Bar{B}_s \to \phi(\to K^+ K^-) \mu^+ \mu^-$  are identical in terms of the final states observed at the experiment,  and hence the flavor of the initial $B_s$-meson can't be determined. Thus, in the absence of tagging, only the $CP$-averaged decay distribution $(d (\Gamma+\bar{\Gamma}))$ can be measured\footnote{See Ref.~\cite{Descotes-Genon:2015hea} for the time-dependent
angular distributions for the decay $B_s \to \phi(\to K^+ K^-) \mu^+ \mu^-$, where the time-integrated observables are suppressed due to $B_s$-meson mixing effects, reducing their experimental sensitivity. Additionally, a few other time-dependent observables are highlighted, which can only be measured via flavor-tagging and are therefore beyond the reach of hadronic colliders.}.
With the expressions of the transversity amplitudes and the angular coefficients given in Appendix~\eqref{app:tAmp}, the differential decay width is given by
\begin{align}
\label{eq:dgamma}
   \frac{d\hs \Gamma_{B_s}}{d\hs q^2} =   \bigg[\frac{3}{4}\left(2 I_1^s+I_1^c\right)-\frac{1}{4}\left(2 I_2^s+I_2^c\right)\bigg] \equiv \Gamma_f\,.
\end{align}
The untagged CP-averaged angular decay distribution gives access to only the following CP-averaged angular observables normalized to the differential decay width defined as
\begin{align}
\label{eq:angular_obs}
    F_L = -\frac{I_{2}^c}{\Gamma_f}\,,\,\,\,
    S_3 = \frac{I_3}{\Gamma_f}\,,\,\,\,
     S_4 = \frac{I_4}{\Gamma_f}\,,\,\,\,
      S_7 = \frac{I_7}{\Gamma_f}\,,\,\,\,
\end{align}
where $F_L$ is the longitudinal polarization
fraction of the $\phi$-meson and $S_{3,4,7}$ are the angular symmetries. In Fig.~\eqref{fig:observables}, we plot these observables highlighting the effects of the non-factorizable charm-loop corrections in comparison to the SM case without these contributions. We also overlay the experimental measurements from Refs. \cite{LHCb:2021zwz,LHCb:2021xxq}.\footnote{Note that the conventions used to define the kinematic angles differ between the theoretical and experimental approaches which result in sign differences for certain observables, such as $S_4$ and $S_7$ \cite{Descotes-Genon:2015uva}.} It can be seen that the uncertainties in the SM prediction of the differential branching fraction are large at high $q^2$ because it is based solely on the LCSR approach, which is less reliable in this region. The uncertainties are reduced while constructing the angular observables which are ratios. The observable $S_7$ is proportional to the imaginary part of the amplitudes (see Eq.~\eqref{eq:I7}), which is tiny in the SM and, therefore, is enhanced only in the resonance regions.  The predictions of the differential branching fraction seem consistent with the experimental data, while the angular observables show deviations in some of the bins, e.g., $S_3$ and $S_7$ in the $[4-6]\,\text{GeV}^2$ bin\footnote{An analysis~\cite{Descotes-Genon:2019bud} on non-resonant $B \to K \pi$ form factors estimates an about $\sim 20\%$ enhancement in the $B \to K^*$ rate due to the effect of a non-vanishing $K^*$ width. Such effect is negligible in $B_s \to \phi$ case, as $\phi$ width is one order smaller than $K^*$ making the contribution less than $\sim 2\%$.}. We refrain from making any remarks about the behavior in the high-$q^2$ regions where Lattice results are more reliable, and also near the charmonium resonances, the predictions are dominated by the long-distance effects.

\section{Summary} \label{sec:summary}

In this analysis, we have investigated the impact of higher-twist (twist-5 and twist-6) three-particle $B_s$-meson LCDAs on the non-local form factors induced by four-quark operators involving charm quarks in the $B_s \to \phi \bar{\ell} \ell$ transition. Our results show an enhancement of approximately one order of magnitude compared to previous studies when the twist-5 and twist-6 contributions are included, with the dominant corrections arising from the twist-5 LCDAs. The observed large values from twist-5 LCDAs are due to the cancellation of LCDA contributions within the same twist and between twist-3 and twist-4. A similar pattern also follows for the $B\rightarrow K^* \bar\ell\, \ell$ decays where the magnitude of the contributions from various LCDAs is slightly smaller compared to $B_s \to \phi \bar{\ell} \ell$. These corrections are highly sensitive to the values of $\lambda_E^2$ and $\lambda_H^2$, which are not well-determined. To assess this sensitivity, we classified the various determinations available in the literature into five categories and found that the estimated non-local form factors can vary significantly depending on the chosen $\lambda_{E,H}^2$ values. In contrast, for the local form factors, the dominant contributions come from the two-particle LCDAs, making the impact of higher-twist three-particle LCDAs less than one percent. 

The local form factors obtained in the spacelike region are analytically continued to the timelike region by using the $z$-expansion, whereas for the non-local amplitudes, double-subtracted dispersion relations are used. In the latter case, the parameters are determined by fitting to the LCSR predictions at the negative $q^2$ values and incorporating data from the two-body decays such as $B_s \to \phi\,\phi$, $B_s\to \phi\,J/\psi$ and $B_s \to \phi\,\psi(2S)$, due to the presence of vector resonances in the kinematically allowed region. This approach allows us to express the total charm-loop contribution as a $q^2$-dependent correction to the Wilson coefficient $C_9^{\text{eff}}$ which varies for each polarization of the final-state $\phi$-meson. These $q^2$-dependent corrections are found to be higher than the SM predictions without the non-factorizable charm-loop corrections but remain consistent within uncertainties. The phenomenological impact of these contributions on various observables of this mode is also analyzed. We find that the predictions of the differential branching fraction are consistent with the experimental data.
\subsubsection*{Acknowledgment}
We thank Alexander Khodjamirian for insightful discussions and Nico Gubernari for his assistance in reproducing results from the literature. A.K. and R.M. acknowledge support from the SERB Grant SPG/2022/001238. I.R.'s work is supported by the NSERC of Canada.

\appendix

\section{Running of Wilson coefficients}\label{app:match1}

\begin{table}[H]
    \centering
\renewcommand*{\arraystretch}{1.1}
    \begin{tabular}{||c|c|c||}\hline\hline
        Description& Parameter& Value  \\ \hline  
          Bottom quark mass &$m_b\equiv M_b$& $4.78\pm0.06\GeV$\\ \hline
          Higgs boson mass &
           $M_H$& $125.20\pm0.11\hs \GeV$\\ \hline
         Top quark mass&  $\overline{m}_t$ & $162.5\pm 2.1\hs \GeV$\\ \hline
         $Z$-Boson mass& $M_Z$& $91.1880 \hs \GeV$\\ \hline
          $W$-Boson mass & $M_W$& $80.3692\hs  \GeV $ \\ \hline 
          Renormalization scale &$\mu_b$& $[M_b/2,2\hs M_b]$ \\ \hline
           Fine-structure constant&$\alpha_e(M_Z)$& $1/127.930$ \\\hline
           Strong coupling constant&
           $\alpha_s(M_Z)$  &$0.1180\pm 0.0009$ \\ \hline\hline
    \end{tabular}
    \caption{Input parameters from PDG\cite{ParticleDataGroup:2022pth}  used to calculate the running of the Wilson coefficients in the weak effective theory.}
    \label{tab:WC_inputs}
\end{table}

The Wilson coefficients $C_i$ of the weak effective field theory are evolved to the bottom quark mass scale using the known NNLO QCD and NLO electroweak results from Refs.~\cite{Bobeth:1999mk,Buchalla:1997kz,Gambino:2001au,Bobeth:2003at,Gorbahn:2004my,Huber:2005ig,Gorbahn:2005sa,Gambino:2003zm} and taking numerical inputs from Table~\eqref{tab:WC_inputs}. For the bottom and charm quarks, the pole mass definition is used while the strange quark mass is taken at $2\hs \GeV$ in the  $\overline{\mathrm{MS}}$ scheme. The renormalization group running of the strong coupling constant is performed using RunDec~\cite{Herren:2017osy}, and the decoupling is performed at the pole mass definitions of the heavy quarks. The effective coefficients $C_i^\text{eff.}$ are defined as
\begin{align}
    C_7^\text{eff.}&=\frac{4\pi}{\alpha_s} C_7-\frac{1}{3}C_3-\frac{4}{9}C_4-\frac{20}{3}C_5-\frac{80}{9}C_6\,,\\ 
    C_8^\text{eff.}&=\frac{4\pi}{\alpha_s} C_8+C_3-\frac{1}{6}C_4+20 C_5-\frac{10}{3}C_6\,,\\
    C_{9,10}^\text{eff.}&=\frac{4\pi}{\alpha_s}C_{9,10}\,,
\end{align}
and their values at different scales are presented in Table~\eqref{tab:wilson_wet}.

\begin{table}[H]
    \centering
\renewcommand*{\arraystretch}{1.1}
    \begin{tabular}{||c|c|c|c|c||}
        \hline\hline
 Wilson coefficient& $\mu=\overline{m}_b$ & $\mu=M_b/2$& $\mu=M_b $& $\mu=2 M_b$\\ \hline \hline
 $C_1(\mu)$& -0.291 & -0.434 & -0.261 & -0.127 \\ \hline
 $C_2(\mu)$& 1.016 & 1.027 & 1.014 & 1.007 \\ \hline
 $C_3(\mu)$& -0.006 & -0.011 & -0.005 & -0.002 \\ \hline
 $C_4(\mu)$& -0.086 & -0.127 & -0.079 & -0.050 \\ \hline
 $C_5(\mu)$& 0.0004 & 0.0008 & 0.0003 & 0.0001 \\ \hline
 $C_6(\mu)$& 0.0011 & 0.0023 & 0.0009& 0.0003 \\ \hline
 $C_7(\mu)$& -0.006 & -0.008 & -0.006 & -0.004 \\ \hline
 $C_8(\mu)$& -0.003 & -0.004 & -0.003 & -0.002 \\ \hline
 $C_9(\mu)$& 0.079 & 0.101 & 0.075 & 0.056 \\ \hline
 $C_{10}(\mu)$& -0.079 & -0.095 & -0.075 & -0.063 \\ \hline
 $\text{C}_7^{\text{eff.}}(\mu)$& -0.304 & -0.326 & -0.299 & -0.275 \\ \hline
 $\text{C}_8^{\text{eff.}}(\mu)$& -0.168 & -0.184 & -0.165 & -0.151 \\ \hline
 $\text{C}_9^{\text{eff.}}(\mu)$& 4.440 & 4.675 & 4.370 & 3.936 \\ \hline
 $\text{C}_{10}^{\text{eff.}}(\mu)$& -4.398 & -4.416 & -4.394 & -4.370 \\ \hline\hline
 \end{tabular}
\caption{Wilson coefficients obtained using $\overline{m}_b=4.183\GeV$, $M_b=4.78\GeV$ and other inputs from Table~\eqref{tab:WC_inputs}.  }
    \label{tab:wilson_wet}
\end{table}

 The factorizable correction to the Wilson coefficient $C_7$  at the leading order in $\alpha_s$ is $\Delta\hs C_{7}^{(0)}=0$ and the NLO corrections are calculated in Ref.~\cite{Asatrian:2019kbk}. The leading order contribution for the $C_9$ i.e.,  $\Delta\hs C_9^{(0)}$ is given by
\begin{align}
    \Delta C_9^{(0)}=\frac{2}{3} Q_c \left(C_F \hs C_1+C_2\right) \Bigg\lbrace &\frac{2}{3}+i \hs \pi+\frac{4z}{s}+\log\left(\frac{4\mu^2}{m_b^2}\right)+2\log(x)-\log(1-x)-\log(1+x)\nonumber\\&+\frac{1-3y^2}{2y^3}\left(\log(1+y)-\log(1-y)\right)\Bigg\rbrace\,,
\end{align}
where,
\begin{equation}
    z=\frac{m_c^2}{m_b^2},\quad s=\frac{q^2}{m_b^2},\quad x=\left(1-4\hs z\right)^{-\frac{1}{2}},\quad y=\left(1-4\hs z/s\right)^{-\frac{1}{2}}\,,
\end{equation}
and the NLO contributions, which we denote by $\Delta C_{7,9}^{\text{NLO}}$, are calculated using the codes provided in Ref.~\cite{Asatrian:2019kbk}.

\section{\texorpdfstring{$B_s$}{}-meson light-cone distribution amplitudes}
\label{app:BDA}
	\subsection{Two-particle distribution amplitudes}
 We express the $B_s$-to-vacuum matrix element in the heavy quark effective theory limit in which the heavy $b$-quark field is substituted with the heavy quark effective theory field $h_v$, with $v^\mu=\left( q+k\right)^\mu/m_{B_s}$ representing the $B_s$-meson's four-velocity in its rest frame.
The expansion in terms of the different twists of two-particle $B_s$-meson LCDAs is given as \cite{Braun:2017liq}
	\begin{align}
	\bra{0} \bar{s}^{a}(x) h_{v}^{b}(0) \ket{\bar{B_s}(v)} =
	-\frac{i f_{B_s} m_{B_s}}{4} \int^\infty_0& d\omega  \bigg\{
	(1 + \slashed{v}) \bigg[
	\phi_+(\omega)
	- g_+(\omega) \partial_\lambda \partial^\lambda\nonumber\\ &+ \frac12 \left(\overline{\phi}(\omega)- \overline{g}(\omega)
	 \partial_\lambda \partial^\lambda\right) 
	\gamma^\rho \partial_\rho
	\bigg] \gamma_5
	\bigg\}^{ba} e^{-i r \cdot x}
	\Bigg|_{r=\omega v}
	\,,
	\label{eq:BLCDAs2pt}
	\end{align}
where \begin{equation}
    \overline{\phi}(\omega)\equiv\int_0^\omega d\eta \left(\phi_+(\eta)-\phi_-(\eta)\right),\text{ and }\quad 
    \overline{g}(\omega)\equiv\int_0^\omega d\eta \left(g_+(\eta))-g_-(\eta))\right)\,.
\end{equation}
 The derivatives in Eq.~\eqref{eq:BLCDAs2pt} defined as $\partial_\mu \equiv \partial/\partial r^\mu$, are understood to be acting on the hard-scattering kernel. The higher-derivative terms i.e., $\mathcal{O}(x^4)$ terms, in the light-cone expansion are neglected in Eq.~\eqref{eq:BLCDAs2pt}. This expansion is adequate for calculating the higher-twist corrections at leading order in the strong coupling constant. \par
 The Exponential model for the two-particle $B_s$-meson LCDAs for different twists is given as \cite{Grozin:1996pq,Braun:2017liq,Lu:2018cfc}
	\begin{align}
	\text{Twist-2:}\quad & \phi_+(\omega)= \frac{\omega}{\omega_0^2}e^{-\omega/\omega_0}\,,\\
	\text{Twist-3:}\quad &\phi_-(\omega)
	= \frac{1}{\omega_0}e^{-\omega/\omega_0} - \frac{\lambda_E^2 - \lambda_H^2}{18 \omega_0^5} \left(2 \omega_0^2 - 4 \omega \omega_0 + \omega^2 \right)e^{-\omega/\omega_0}\nonumber\,,\\
	\text{Twist-4:}\quad& g_+(\omega)=
	-\frac{\lambda_E^2}{6\omega_0^2}
	\biggl\{(\omega -2 \omega_0)  \text{Ei}\left(-\frac{\omega}{\omega_0}\right) +  
	(\omega +2\omega_0) e^{-\omega/\omega_0}\left(\ln \frac{\omega}{\omega_0}+\gamma_E\right)-2 \omega e^{-\omega/\omega_0}\biggr\} \nn \\ 
 &\bs\bs+ \frac{e^{-\omega/\omega_0}}{2{ \omega_0}}\omega^2\biggl\{1 - \frac{1}{36\omega_0^2}(\lambda_E^2- \lambda_H^2)\biggr\}\nonumber, \label{gplus-model1} \\
 \text{Twist-5:}\quad & g_-(\omega)=\omega \Bigg\lbrace\frac{3}{4}-\frac{\Le{E}-\Le{H}}{12\omega_0^2}\left[1-\frac{\omega}{\omega_0}+\frac{1}{3}\left(\frac{\omega}{\omega_0}\right)^2\right]\Bigg \rbrace e^{-\omega/\omega_0}\,,
 \end{align}
where $\omega_0=\lambda_{B_s}$ and Ei(x) is the Exponential integral. It should be noted that the form of $g_-(\omega)$ also depends on the three particle twist-5 LCDA $\psi_5(\omega_1,\omega_2)$ due to QCD EOM~\cite{Braun:2017liq, Lu:2018cfc}.
 	\subsection{Three-particle distribution amplitudes}
  \label{app:3_LCDA}

The gluon contributions to the quark propagator result in the three-particle contributions to the matrix element in Eq.~\eqref{eq:BLCDAs2pt} which are given by
\begin{align}
    \bra{0} &\bar{s}^a(x) 
G_{\sigma\tau}(u x) h_v^b(0) \ket{B_s(v)}
    \nonumber\\
    & =
    \frac{f_{B_s} m_{B_s}}{4} \int_0^\infty d\omega_1 \int_0^\infty d\omega_2 e^{-i (\omega_1+u \omega_2) v\cdot x}\,    
    \bigg\{(1+\slashed{v})\bigg[
        (v_\sigma\gamma_\tau-v_\tau\gamma_\sigma)[\psi_A-\psi_V] 
        -i\sigma_{\sigma\tau}\psi_V  
    \nonumber\\
    & \bs\bs\bs+ (\partial_\sigma v_\tau-\partial_\tau v_\sigma) \overline{X}_A
      - (\partial_\sigma\gamma_\tau-\partial_\tau\gamma_\sigma) [\overline{W} + \overline{Y}_A]
      + i\epsilon_{\sigma\tau\alpha\beta}\partial^\alpha v^\beta\gamma_5 \overline{\tilde{X}}_A 
    \nonumber\\
    &\bs\bs\bs - i\epsilon_{\sigma\tau\alpha\beta}\partial^\alpha \gamma^\beta\gamma_5 \overline{\tilde{Y}}_A
      - u (\partial_\sigma v_\tau-\partial_\tau v_\sigma)\slashed{\partial} \overline{\overline{W}}
      + u(\partial_\sigma \gamma_\tau-\partial_\tau \gamma_\sigma)\slashed{\partial} \overline{\overline{Z}}
    \bigg]\gamma_5\bigg\}^{ba}
    \,,\label{eq:MEL3}
\end{align}
 where the $B_s$-meson distribution amplitudes ($\psi_A,\psi_V,X_A,\tilde{X}_A, W,Y_A,\tilde{Y}_A,\mathrm{and}\, Z$) are the functions of $\omega_1$ and $\omega_2$ and the derivative $\partial_\mu \equiv \partial/\partial (v^\mu(\omega_1+u \omega_2))$ acts on the hard kernel. The barred DAs are obtained using:
\begin{align}
\overline{X}(\omega_1,\omega_2)&\equiv \int_0^{\omega_1} d\eta_1 X(\eta_1,\omega_2)\,,\\
\overline{\overline{X}}(\omega_1,\omega_2)&\equiv \int_0^{\omega_1} d\eta_1 \int_0^{\omega_2}d\eta_2 X(\eta_1,\eta_2)\,.
\end{align}
Now in the case of non-local form factor calculation, the soft-gluon corrections to the four-quark operator give rise to the effective non-local quark-antiquark-gluon operator. This can be
obtained from the three-particle contributions in Eq.~\eqref{eq:MEL3} by writing the gluon field strength tensor using the non-local translation operator i.e. $G_{\mu\nu}(u x)=e^{-iu x\cdot (i\,\mathcal{D})}G_{\mu\nu}(0)$ in terms of the covariant derivative operator ($\mathcal{D}$) in the fixed-point gauge. The covariant derivative operator can be decomposed in terms of the light-cone vectors, $n_{+/-}$, and the large contributions from the gluon emitted along the direction of $s$-quark are then obtained from the $n_-$ component of this operator. Under these approximations, we can write
\begin{align}
    G_{\mu\nu}( u\, x)\simeq e^{-i\, u \frac{(n_-\cdot\, x)}{2}(i\,n_+ \cdot\, \mathcal{D})}G_{\mu\nu}(0)=\int d \omega_2 \, e^{-i u\,\frac{(n_-\cdot\,x) }{2}\omega_2}\delta\left[\omega_2-i n_+\cdot\,\mathcal{D}\right] G_{\mu\nu}(0)\,,
    \label{eq:gluon_sum}
\end{align}
and the non-local $B_s$-to-vacuum matrix element is obtained using Eq.~\eqref{eq:gluon_sum} in Eq.~\eqref{eq:MEL3} as
\begin{align}
    \bra{0} &\bar{s}^a(x) \delta \left[ \omega_2 - i n_+ \cdot \mathcal{D} \right]
G_{\sigma\tau} h_v^b(0) \ket{B_s(v)}\Big|_{x\simeq n_+, x^2\simeq 0}
    \nonumber\\
    & =
    \frac{f_{B_s} m_{B_s}}{4} \int_0^\infty d\omega_1 e^{-i \omega_1 v\cdot x}\,    
    \bigg\{(1+\slashed{v})\bigg[
        (v_\sigma\gamma_\tau-v_\tau\gamma_\sigma)[\psi_A-\psi_V] 
        -i\sigma_{\sigma\tau}\psi_V  
    \nonumber\\
    & \bs\bs\bs+ (\partial_\sigma v_\tau-\partial_\tau v_\sigma) \overline{X}_A
      - (\partial_\sigma\gamma_\tau-\partial_\tau\gamma_\sigma) [\overline{W} + \overline{Y}_A]
      + i\epsilon_{\sigma\tau\alpha\beta}\partial^\alpha v^\beta\gamma_5 \overline{\tilde{X}}_A 
    \nonumber\\
    &\bs\bs\bs - i\epsilon_{\sigma\tau\alpha\beta}\partial^\alpha \gamma^\beta\gamma_5 \overline{\tilde{Y}}_A
      - (\partial_\sigma v_\tau-\partial_\tau v_\sigma)\slashed{\partial} \overline{\overline{W}}
      + (\partial_\sigma \gamma_\tau-\partial_\tau \gamma_\sigma)\slashed{\partial} \overline{\overline{Z}}
    \bigg]\gamma_5\bigg\}^{ba}
    \,.\label{eq:ME3}
\end{align}
Here, note a resulting modification in the definitions of derivative ($\partial_\mu \equiv \partial/\partial (v^\mu\omega_1)$) and the double-barred distribution amplitudes compared to the previous case (Eq.~\eqref{eq:MEL3}) which is given by
\begin{align}
\overline{\overline{X}}(\omega_1,\omega_2)&\equiv \int_0^{\omega_1} d\eta_1 \int_0^{\eta_1}d\eta_2 X(\eta_2,\omega_2)\,.
\end{align}

In order to rewrite Eqs.~\eqref{eq:MEL3} and \eqref{eq:ME3} in terms of their twist expansion, the distribution amplitudes can be recast as the LCDAs of a definite twist as
\begin{equation}
 \begin{aligned}
    \psi_A & = 
    \frac{1}{2}\, [ \phi_3 + \phi_4                                             ] 
    \,, \quad \hspace{.3cm}
    X_A  = 
    \frac{1}{2}\, [-\phi_3 - \phi_4 + 2\psi_4                                   ] 
    \,, \quad 
    Y_A = 
    \frac{1}{2}\, [-\phi_3 - \phi_4 +  \psi_4 - \psi_5                          ] 
    \,, \\
    \psi_V  &= 
    \frac{1}{2}\, [-\phi_3 + \phi_4                                             ] 
    \,,\quad
    \tilde{X}_A = 
    \frac{1}{2}\, [-\phi_3 + \phi_4 - 2\tilde{\psi}_4                                   ] 
    \,, \quad 
    \tilde{Y}_A  = 
    \frac{1}{2}\, [-\phi_3 + \phi_4 -  \tilde{\psi}_4 + \tilde{\psi}_5                          ]
    \,, \\
    W           & = 
    \frac{1}{2}\, [ \phi_4 -\psi_4 -  \tilde{\psi}_4 + \tilde{\phi}_5 + \psi_5 + \tilde{\psi}_5] \,, \quad
    Z         = 
    \frac{1}{4}\, [-\phi_3 + \phi_4 - 2\tilde{\psi}_4 + \tilde{\phi}_5 + 2 \tilde{\psi}_5 - \phi_6]\,,
\end{aligned}
\label{eq:DA_twist_expansion}
\end{equation}
where $\phi_3$ and $\phi_4 $,$\psi_4$,$\tilde{\psi}_4$ are the LCDAs of twist-3 and twist-4, $\psi_5 $, $\tilde{\psi}_5 $, $\tilde{\phi}_5 $ and $\phi_6$ are the LCDAs of twist-5 and twist-6, respectively. For the numerical analysis in this paper, we use the Exponential Model in which the three-particle $B_s$-meson LCDAs are parametrized as~\cite{Braun:2017liq, Lu:2018cfc}
\begin{equation}
\begin{aligned}
\label{eq:3_LCDA_Expo}
    \phi_3(\omega_1,\omega_2)&=\frac{\Le{E}-\Le{H}}{6 \omega_0^5}\omega_1\omega_2^2\,e^{-\left(\omega_1+\omega_2\right)/\omega_0}\,,&
    \phi_4(\omega_1,\omega_2)&=\frac{\Le{E}+\Le{H}}{6 \omega_0^4}\omega_2^2\,e^{-\left(\omega_1+\omega_2\right)/\omega_0}\,, \\
    \psi_4(\omega_1,\omega_2)&=\frac{\Le{E}}{3 \omega_0^4}\omega_1\omega_2\,e^{-\left(\omega_1+\omega_2\right)/\omega_0}\,,&        \tilde{\psi}_4(\omega_1,\omega_2)&=\frac{\Le{H}}{3 \omega_0^4}\omega_1\omega_2\,e^{-\left(\omega_1+\omega_2\right)/\omega_0}\,,\\
    \tilde{\phi}_5(\omega_1,\omega_2)&=\frac{\Le{E}+\Le{H}}{3 \omega_0^3}\omega_1\,e^{-\left(\omega_1+\omega_2\right)/\omega_0}\,,&  \psi_5(\omega_1,\omega_2)&=-\frac{\Le{E}}{3\omega_0^3}\omega_2\,e^{-\left(\omega_1+\omega_2\right)/\omega_0}\,, \\
    \tilde{\psi}_5(\omega_1,\omega_2)&=-\frac{\Le{H}}{3 \omega_0^3}\omega_2\,e^{-\left(\omega_1+\omega_2\right)/\omega_0}\,,& \phi_6(\omega_1,\omega_2)&=\frac{\Le{E}-\Le{H}}{3 \omega_0^2}\,e^{-\left(\omega_1+\omega_2\right)/\omega_0}\,.
\end{aligned}
\end{equation}

\section{Hadronic matrix elements in the transversity basis}
\label{app:had}
In this Appendix, we define the hadronic matrix elements for the local and non-local form factors used in the transversity basis and present the transformation relations from the traditional basis.  The transversity basis is associated with specific Lorentz decompositions of the matrix elements with the following structures
\begin{equation}
\begin{aligned}
    \label{eq:def:Lor}
    \mathcal{L}_{\alpha\mu}^{\perp}(k, q) & = \frac{\sqrt{2}\, m_{B_s}}{\sqrt{\tilde{\lambda}_{q^2}}} \epsilon_{\alpha\mu k q}\,, &
    \mathcal{L}_{\alpha\mu}^{\parallel}(k, q) & = \frac{i\, m_{B_s}}{\sqrt{2}\tilde{\lambda}_{q^2}}\left(g_{\alpha\mu}\hs \tilde{\lambda}_{q^2} - 4(q\cdot k) q_\alpha k_\mu
                          + 4 m_{\phi}^2  \,q_\alpha q_\mu\right)\,, \\
    \mathcal{L}_{\alpha\mu}^{t}(k, q)     & = \frac{2i\, m_{\phi}}{q^2} q_\alpha q_\mu\,, &
    \mathcal{L}_{\alpha\mu}^{0}(k, q)     & = \frac{4i\, m_{\phi} m_{B_s}^2}{q^2 \tilde{\lambda}_{q^2}}
                            \left(q^2 q_\alpha k_\mu - (q\cdot k)\,q_\alpha q_\mu\right)\,.
\end{aligned}   
\end{equation} 

Using the above decomposition, the local $B_s \to \phi $ matrix elements for the vector (tensor) currents given in Eqs.~\eqref{eq:SLFF1} -- \eqref{eq:mat_el_T1} can be written in terms of redefined form factors $\mathcal{F}_{\lambda}(\mathcal{F}_{\lambda,T})$ as
\begin{align}
\bra{\phi(k)} \bar{s}\gamma_\mu P_L \,b \ket{B_s(q+k)}
     &=\frac{1}{2}\, \epsilon^{*\alpha} \left[ \mathcal{L}_{\alpha\mu}^{\perp} \mathcal{F}_{\perp}
        -  \mathcal{L}_{\alpha\mu}^{\parallel} \mathcal{F}_{\parallel} - \mathcal{L}_{\alpha\mu}^{0} \mathcal{F}_{0} - \mathcal{L}_{\alpha\mu}^{t} \mathcal{F}_{t}\right]
    \,,\\
\bra{\phi(k)} \bar{s}\sigma_{\mu\nu} q^\nu P_R\, b\ket{B_s(q+k)}
   &= \frac{i}{2}\, m_{B_s}\, \epsilon^{*\alpha}\, \left[\mathcal{L}_{\alpha\mu}^{\perp} \mathcal{F}_{\perp,T}
    -\mathcal{L}_{\alpha\mu}^{\parallel} \mathcal{F}_{\parallel,T} - \mathcal{L}_{\alpha\mu}^{0} \mathcal{F}_{0,T}\right]\,,
\end{align}
which are related to the traditional basis of form factors $V,\,A_i,\,T_i$ as \cite{Gubernari:2020eft,Khodjamirian:2006st}
\begin{equation}
\begin{aligned}
    &\mathcal{F}_{\perp}         = \frac{\sqrt{2}\tilde{\lambda}_{q^2}^{1/2}}{m_{B_s} (m_{B_s} + m_{\phi})} V\,, &
    &\mathcal{F}_{\parallel}         = \frac{\sqrt{2}\,(m_{B_s} + m_{\phi})}{m_{B_s}} A_1\,,       \\
    &\mathcal{F}_{0}             = \frac{(m_{B_s}^2 - m_{\phi}^2 - q^2)(m_{B_s} + m_{\phi})^2 A_1 - \tilde{\lambda}_{q^2}\, A_2}{2 m_{\phi} m_{B_s}^2 (m_{B_s} + m_{\phi}) }\,, &
    &\mathcal{F}_{t}             = A_0\,,\\
    &\mathcal{F}_{\perp,T}   = \frac{\sqrt{2}\,\tilde{\lambda}_{q^2}^{1/2}}{m_{B_s}^2} T_1\,, &
   & \mathcal{F}_{\parallel,T}  = \frac{\sqrt{2} (m_{B_s}^2 - m_{\phi}^2)}{m_{B_s}^2} T_2\,, \\
    & \mathcal{F}_{0,T}      = \frac{q^2 (m_{B_s}^2 + 3 m_{\phi}^2 - q^2)}{2 m_{B_s}^3 m_{\phi}} T_2
    -\frac{q^2\tilde{\lambda}_{q^2}}{2 m_{B_s}^3 m_{\phi} (m_{B_s}^2 - m_{\phi}^2)} T_3\,.\label{eq:ff_convert}
\end{aligned}   
\end{equation}

A similar decomposition can also be performed for the non-local matrix element appearing in Eq.~\eqref{eq:H_mu_NL} which is shown in Eq.~\eqref{eq:Hmu_NLlam}. In this case, the non-local amplitudes in the transversity basis are expressed in terms of the non-local form factors $\Vff{\lambda}$ as
\begin{align}
\widetilde{\mathcal{H}}_\lambda = 2 Q_c \left(C_2 - \frac{C_1}{2 N_c} \right) \Vff{\lambda}\,,
\end{align}
which are related to the traditional basis non-local form factors $\Vff{i}$ as \cite{Gubernari:2020eft}
\begin{equation}
    \label{eq:rel:V_i-to-us}
    \begin{aligned}
    \mathcal{\widetilde{V}}_{\perp}  & =  \frac{\tilde{\lambda}^{1/2}_{q^2}}{\sqrt{2} m_{B_s}^3} \, \mathcal{\widetilde{V}}_{1}\,,       \\
    \mathcal{\widetilde{V}}_{\parallel} & = -\sqrt{2}\frac{m_{B_s}^2 - m_{\phi}^2}{m_{B_s}^3} \,  \mathcal{\widetilde{V}}_{2}\,, \\
    \mathcal{\widetilde{V}}_{0}     & = -
   \frac{q^2}{2 m_{B_s}^4 m_{\phi}}\, 
     \left[(m_{B_s}^2 + 3 m_{\phi}^2 - q^2)\mathcal{\widetilde{V}}_{2} - \frac{\tilde{\lambda}_{q^2}}{m_{B_s}^2 - m_{\phi}^2} \mathcal{\widetilde{V}}_{3} \right]\,.
    \end{aligned}
\end{equation}

\begin{table}[ht]
\begin{adjustbox}{max width=\textwidth}
    \centering
    \renewcommand*{\arraystretch}{1.3}
    \begin{tabular}{||c||c||c||c|c|c||c|c|c||c||c||}
    \hline\hline
    \multirow{2}{2em}{Order} & 
    \multirow{2}{8em}{\centering Traditional DAs} & 
    \multicolumn{8}{c||}{LCDA contributions $(10^{-8})$}&\multirow{2}{2em}{Total}\\ \cline{3-10}
    & &  Twist-3& \multicolumn{3}{c||}{Twist-4}& \multicolumn{3}{c||}{Twist-5}& twist-6&   \\ \hline \hline
 \multirow{2}{4em}{$\mathcal{O}\left(\frac{x}{x\cdot v}\right)^0$} & $\psi _A-\psi _V $ & $\cellcolor{red!25} 54.01_{\phi _3} $ &  &  &  &  &  &  & & 54.01 \\  & $
 \psi _V $ & $\cellcolor{red!25} -36.35_{\phi _3} $ & $\cellcolor{green!25} -452.65_{\phi _4} $ &  &  &  &  &  & & -489.00 \\
 \hline\hline 
\multicolumn{2}{||c||}{Total at $\mathcal{O}\left(\frac{x}{x\cdot v}\right)^0 $}&17.66& -452.65& & & & & & &  -434.99\\ \hline 
\multicolumn{2}{||c||}{Total from twists at $\mathcal{O}\left(\frac{x}{x\cdot v}\right)^0$ } & 17.66& \multicolumn{3}{c||}{-452.65}& \multicolumn{3}{c||}{}& & \\ \hline\hline
 \multirow{4}{4em}{$\mathcal{O}\left(\frac{x}{x\cdot v}\right)^1$}  
 & $
 \overline{X}_A $ & $\cellcolor{red!25} -3.20_{\overline{\phi }_3} $ & $\cellcolor{green!25} 38.50_{\overline{\phi }_4} $ & $\cellcolor{blue!25} -5.19_{\overline{\psi }_4} $ & & &  &  &  &30.11 \\
& $
 \overline{Y}_A+\overline{W} $ & $\cellcolor{red!25} -32.75_{\overline{\phi }_3} $&& & $\cellcolor{blue!25} 77.62_{\overline{\tilde{\psi }}_4} $& & $ \cellcolor{green!25} -139.73_{\overline{\tilde{\phi }}_5} $ & $\cellcolor{blue!25} 329.61_{\overline{\tilde{\psi }}_5} $ & & 234.75    \\  & $
 \overline{\tilde{X}}_A $ & $\cellcolor{red!25} 0.06_{\overline{\phi }_3} $ & $ \cellcolor{green!25} 0.84_{\overline{\phi }_4} $ & & $\cellcolor{blue!25} -0.65_{\overline{\tilde{\psi }}_4} $ &  &  & & & 0.25 \\  & $
 \overline{\tilde{Y}}_A $ & $\cellcolor{red!25} 34.04_{\overline{\phi }_3} $ & $ \cellcolor{green!25} 425.00_{\overline{\phi }_4} $& & $\cellcolor{blue!25} -80.60_{\overline{\tilde{\psi }}_4} $ & && $\cellcolor{blue!25} -342.10_{\overline{\tilde{\psi }}_5} $ &  &36.34 \\ \hline\hline
 \multicolumn{2}{||c||}{Total at $\mathcal{O}\left(\frac{x}{x\cdot v}\right)^1\,$} & -1.85& 464.34&  -5.19 &-3.63& &-139.73& 12.49& & 301.45\\ 
 \hline
 \multicolumn{2}{||c||}{Total from twists at $\mathcal{O}\left(\frac{x}{x\cdot v}\right)^1$ } & 1.85& \multicolumn{3}{c||}{455.52}& \multicolumn{3}{c||}{-152.22}& & \\ \hline\hline
 \multirow{2}{4em}{$\mathcal{O}\left(\frac{x}{x\cdot v}\right)^2$} & $
 \overline{\overline{W}} $& & $\cellcolor{green!25} -1.48_{\overline{\overline{\phi }}_4} $ & $\cellcolor{blue!25} 0.10_{\overline{\overline{\psi }}_4} $ & $\cellcolor{blue!25} 0.20_{\overline{\overline{\tilde{\psi }}}_4} $ & $\cellcolor{blue!25} 0.40_{\overline{\overline{\psi }}_5} $ & $ \cellcolor{green!25} -0.24_{\overline{\overline{\tilde{\phi }}}_5} $ & $\cellcolor{blue!25} 0.81_{\overline{\overline{\tilde{\psi }}}_5} $ &  & -0.21\\  & $
 \overline{\overline{Z}} $ & $\cellcolor{red!25} -0.58_{\overline{\overline{\phi }}_3} $ & $\cellcolor{green!25} -7.12_{\overline{\overline{\phi }}_4} $& & $\cellcolor{blue!25} 2.76_{\overline{\overline{\tilde{\psi }}}_4} $& & $ \cellcolor{green!25} -2.50_{\overline{\overline{\tilde{\phi }}}_5} $ & $\cellcolor{blue!25} 11.58_{\overline{\overline{\tilde{\psi }}}_5} $ & $\cellcolor{red!25} -3.58_{\overline{\overline{\phi }}_6} $ & 0.56 \\ \hline\hline
 \multicolumn{2}{||c||}{Total at $\mathcal{O}\left(\frac{x}{x\cdot v}\right)^2 $}& -0.58& -8.6& 0.1& 2.96& 0.4& -2.74& 12.39& -3.58& 0.35\\ \hline
 \multicolumn{2}{||c||}{Total from twists at $\mathcal{O}\left(\frac{x}{x\cdot v}\right)^2$ } & -0.58& \multicolumn{3}{c||}{-5.54}& \multicolumn{3}{c||}{10.05}&-3.58 & \\ \hline \hline
 \multicolumn{2}{||c||}{Total from individual LCDA } & 15.23& 3.09& -5.09& -0.67& 0.4&-142.47& -0.1& -3.58& -133.19\\ \hline
 \multicolumn{2}{||c||}{Total from definite twists} & 15.23& \multicolumn{3}{c||}{-2.67}& \multicolumn{3}{c||}{-142.17}& -3.58&  -133.19\\ \hline\hline
 \end{tabular}
    \end{adjustbox}
    \caption{Contribution to the non-local form factor $\Vff{\|}$ at $q^2=-1\,\GeV^2$ arising from various DAs with definite twists and DAs associated with different Lorentz structures. The DAs parametrized in Exponential Model highlighted in green are $\propto |\lambda_E^2+\lambda_H^2|$, in blue are $\propto |\lambda_{E/H}^2|$ and in red are $\propto |\lambda_E^2-\lambda_H^2|$.}
    \label{tab:FFNL_par}
\end{table}

\section{A break up of LCDA contributions to non-local form factors}\label{app:LCDA_contributions}

In this Appendix, we provide the 
detailed breakup of contributions from different LCDAs to the non-local form factors $\Vff{\parallel}$ and $\Vff{0}$ at $q^2=-1\,\gev^2$, similar to the discussion for $\Vff{\perp}$ given in Sec.~\eqref{sec:num_nonlocal}. Here, the results 
are provided in Table~\eqref{tab:FFNL_par} and~\eqref{tab:FFNL_0}, respectively. 

\begin{table}[H]
\begin{adjustbox}{max width=\textwidth}
    \centering
    \renewcommand*{\arraystretch}{1.3}
    \begin{tabular}{||c||c||c||c|c|c||c|c|c||c||c||}
    \hline\hline
    \multirow{2}{2em}{Order} & 
    \multirow{2}{8em}{\centering Traditional DAs} & 
    \multicolumn{8}{c||}{LCDA contributions $(10^{-8})$}&\multirow{2}{2em}{Total}\\ \cline{3-10}
    & &  Twist-3& \multicolumn{3}{c||}{Twist-4}& \multicolumn{3}{c||}{Twist-5}& twist-6&   \\ \hline \hline
 \multirow{2}{4em}{$\mathcal{O}\left(\frac{x}{x\cdot v}\right)^0$}& $\psi _A-\psi _V $ & $\cellcolor{red!25} 0.19_{\phi _3} $ &  &  &  &  &  &  & & 0.19 \\  & $
 \psi _V $ & $\cellcolor{red!25} 0.78_{\phi _3} $ & $\cellcolor{green!25} 9.26_{\phi _4} $ &  &  &  &  &  &  & 10.04\\
 \hline\hline
 \multicolumn{2}{||c||}{Total at $\mathcal{O}\left(\frac{x}{x\cdot v}\right)^0 $}& 0.97& 9.26& & & & & & & 10.23\\ \hline
  \multicolumn{2}{||c||}{Total from twists at $\mathcal{O}\left(\frac{x}{x\cdot v}\right)^0$ } & 0.97& \multicolumn{3}{c||}{9.26}& \multicolumn{3}{c||}{}& & \\ \hline\hline
 \multirow{4}{4em}{$\mathcal{O}\left(\frac{x}{x\cdot v}\right)^1$}
  & $
 \overline{X}_A $ & $\cellcolor{red!25} 0.13_{\overline{\phi }_3} $ & $ \cellcolor{green!25} -2.02_{\overline{\phi }_4} $ & $\cellcolor{blue!25} 0.28_{\overline{\psi }_4} $ & &  &  &  & & -1.61  \\
 & $
 \overline{Y}_A+\overline{W} $ & $\cellcolor{red!25} -0.91_{\overline{\phi }_3} $& & & $\cellcolor{blue!25} 2.02_{\overline{\tilde{\psi }}_4} $ & & $ \cellcolor{green!25} -3.41_{\overline{\tilde{\phi }}_5} $ &\cellcolor{blue!25} $ 6.71_{\overline{\tilde{\psi }}_5} $   &  & 4.41\\  & $
 \overline{\tilde{X}}_A $ & $\cellcolor{red!25} 0.30_{\overline{\phi }_3} $ & $ \cellcolor{green!25} 2.14_{\overline{\phi }_4} $ & & $\cellcolor{blue!25} -1.33_{\overline{\tilde{\psi }}_4} $ &  &  &    & & 1.11\\  & $
 \overline{\tilde{Y}}_A $ & $\cellcolor{red!25} -1.58_{\overline{\phi }_3} $ & $\cellcolor{green!25} -15.83_{\overline{\phi }_4} $& & $\cellcolor{blue!25} 3.50_{\overline{\tilde{\psi }}_4} $  & & &$\cellcolor{blue!25} 11.92_{\overline{\tilde{\psi }}_5} $   & &-1.99\\
 \hline\hline
 \multicolumn{2}{||c||}{Total at $\mathcal{O}\left(\frac{x}{x\cdot v}\right)^1 $} & -2.06& -15.71&0.28 & 4.19& &-3.41& 18.63&  & 1.92\\
 \hline
 \multicolumn{2}{||c||}{Total from twists at $\mathcal{O}\left(\frac{x}{x\cdot v}\right)^1$ } & -2.06& \multicolumn{3}{c||}{-11.24}& \multicolumn{3}{c||}{15.22}& & \\ \hline\hline
 \multirow{3}{4em}{$\mathcal{O}\left(\frac{x}{x\cdot v}\right)^2$} & $
 \overline{\overline{W}} $ & & $\cellcolor{green!25} 0.07_{\overline{\overline{\phi }}_4} $ & $\cellcolor{blue!25} -0.01_{\overline{\overline{\psi }}_4} $ & $\cellcolor{blue!25} -0.01_{\overline{\overline{\tilde{\psi }}}_4} $  & $\cellcolor{blue!25} -0.01_{\overline{\overline{\psi }}_5} $ & $ \cellcolor{green!25} 0.02_{\overline{\overline{\tilde{\phi }}}_5} $ & $\cellcolor{blue!25} -0.05_{\overline{\overline{\tilde{\psi }}}_5} $  & & 0.01\\  & $
 \overline{\overline{Z}} $ & $\cellcolor{red!25} 0.62_{\overline{\overline{\phi }}_3} $  & $\cellcolor{green!25} 6.10_{\overline{\overline{\phi }}_4} $ & & $\cellcolor{blue!25} -2.70_{\overline{\overline{\tilde{\psi }}}_4} $& & $ \cellcolor{green!25} 2.21_{\overline{\overline{\tilde{\phi }}}_5} $  & $\cellcolor{blue!25} -8.95_{\overline{\overline{\tilde{\psi }}}_5} $ & $\cellcolor{red!25} 2.49_{\overline{\overline{\phi }}_6} $&-0.23 \\ \hline\hline
 \multicolumn{2}{||c||}{Total at $\mathcal{O}\left(\frac{x}{x\cdot v}\right)^2 $}& 0.62& 6.17& -0.01& -2.71& -0.01& 2.23& -9.00& 2.49& -0.22\\
 \hline
 \multicolumn{2}{||c||}{Total from twists at $\mathcal{O}\left(\frac{x}{x\cdot v}\right)^2$ } & 0.62& \multicolumn{3}{c||}{3.45}& \multicolumn{3}{c||}{-6.78}& 2.49& \\ \hline \hline
 \multicolumn{2}{||c||}{Total from individual LCDA}&-0.47& -0.28& 0.27& 1.48& -0.01& -1.18& 9.63& 2.49& \\ \hline
  \multicolumn{2}{||c||}{Total from definite twists} & -0.47& \multicolumn{3}{c||}{1.47}& \multicolumn{3}{c||}{8.44}& 2.49&  11.93\\ \hline\hline
 \end{tabular}
    \end{adjustbox}
    \caption{Contribution to the non-local form factor $\Vff{0}$ at $q^2=-1\,\GeV^2$. The color code is same as of Table~\eqref{tab:FFNL_par}.}
    \label{tab:FFNL_0}
\end{table}

\section{Fitted correlations for the local form factors}\label{app:num_local_FF}
The correlation in the coefficients of the second order $z$-expansion for the local form factors (see Eq.~\eqref{eq:local_zfit}) in the traditional basis is provided in Table~\eqref{tab:correlation}. One can easily calculate the correlation matrix in the transversity basis using relations between the two bases as quoted in Eq. \eqref{eq:ff_convert}.

\begin{table}[H]
    \centering
     \renewcommand*{\arraystretch}{1.3}
     \resizebox{\textwidth}{!}{
            \begin{tabular}{||c||c|c|c|c|c|c|c|c|c|c|c|c|c|c|c|c|c|c|c|c|c||}
                \hline \hline
 & $a_0^{V}$ & $a_1^{V}$ & $a_2^{V}$ & $a_0^{A_1}$ & $a_1^{A_1}$ & $a_2^{A_1}$ & $a_0^{A_2}$ & $a_1^{A_2}$ & $a_2^{A_2}$ & $a_0^{T_1}$ & $a_1^{T_1}$ & $a_2^{T_1}$ & $a_0^{T_2}$ & $a_1^{T_2}$ & $a_2^{T_2}$ & $a_0^{T_3}$ & $a_1^{T_3}$ & $a_2^{T_3}$ &$a_0^{A_0}$ &$a_1^{A_0}$ &$a_2^{A_0}$ \\ \hline \hline\hline
 $a_0^{V}$ & 1.00 & -0.96 & 0.16 & 1.00 & 0.98 & -0.99 & 1.00 & -0.98 & -0.55 & 1.00 & -0.95 & -0.31 & 1.00 & 0.98 & -0.99 & 1.00 & -0.99 & 0.43 & 0.98 & -0.81 & 0.18 \\ \hline
 $a_1^{V}$ & -0.96 & 1.00 & -0.44 & -0.95 & -0.88 & 0.97 & -0.94 & 0.97 & 0.48 & -0.96 & 1.00 & 0.02 & -0.96 & -0.88 & 0.96 & -0.94 & 0.98 & -0.55 & -0.98 & 0.93 & -0.43 \\ \hline
 $a_2^{V}$ & 0.16 & -0.44 & 1.00 & 0.15 & -0.04 & -0.22 & 0.12 & -0.24 & 0.03 & 0.16 & -0.46 & 0.89 & 0.16 & -0.03 & -0.21 & 0.12 & -0.26 & 0.54 & 0.31 & -0.65 & 0.91 \\ \hline
 $a_0^{A_1}$ & 1.00 & -0.95 & 0.15 & 1.00 & 0.98 & -0.99 & 1.00 & -0.98 & -0.55 & 1.00 & -0.95 & -0.31 & 1.00 & 0.98 & -0.99 & 1.00 & -0.99 & 0.42 & 0.98 & -0.81 & 0.18 \\ \hline
 $a_1^{A_1}$ & 0.98 & -0.88 & -0.04 & 0.98 & 1.00 & -0.96 & 0.99 & -0.94 & -0.57 & 0.98 & -0.86 & -0.50 & 0.98 & 1.00 & -0.95 & 0.99 & -0.94 & 0.33 & 0.92 & -0.69 & 0. \\ \hline
 $a_2^{A_1}$ & -0.99 & 0.97 & -0.22 & -0.99 & -0.96 & 1.00 & -0.99 & 0.98 & 0.52 & -0.99 & 0.96 & 0.25 & -0.99 & -0.96 & 0.99 & -0.99 & 0.99 & -0.48 & -0.98 & 0.83 & -0.21 \\ \hline
 $a_0^{A_2}$ & 1.00 & -0.94 & 0.12 & 1.00 & 0.99 & -0.99 & 1.00 & -0.98 & -0.53 & 1.00 & -0.93 & -0.35 & 1.00 & 0.99 & -0.98 & 1.00 & -0.98 & 0.42 & 0.96 & -0.78 & 0.14 \\ \hline
 $a_1^{A_2}$ & -0.98 & 0.97 & -0.24 & -0.98 & -0.94 & 0.98 & -0.98 & 1.00 & 0.39 & -0.98 & 0.96 & 0.22 & -0.98 & -0.95 & 0.99 & -0.98 & 1.00 & -0.55 & -0.95 & 0.81 & -0.20 \\ \hline
 $a_2^{A_2}$ & -0.55 & 0.48 & 0.03 & -0.55 & -0.57 & 0.52 & -0.53 & 0.39 & 1.00 & -0.55 & 0.46 & 0.30 & -0.55 & -0.57 & 0.49 & -0.53 & 0.44 & 0.35 & -0.61 & 0.55 & -0.26 \\ \hline
 $a_0^{T_1}$ & 1.00 & -0.96 & 0.16 & 1.00 & 0.98 & -0.99 & 1.00 & -0.98 & -0.55 & 1.00 & -0.95 & -0.31 & 1.00 & 0.98 & -0.99 & 1.00 & -0.99 & 0.42 & 0.97 & -0.81 & 0.18 \\ \hline
 $a_1^{T_1}$ & -0.95 & 1.00 & -0.46 & -0.95 & -0.86 & 0.96 & -0.93 & 0.96 & 0.46 & -0.95 & 1.00 & 0.00 & -0.95 & -0.87 & 0.96 & -0.94 & 0.97 & -0.55 & -0.98 & 0.94 & -0.45 \\ \hline
 $a_2^{T_1}$ & -0.31 & 0.02 & 0.89 & -0.31 & -0.50 & 0.25 & -0.35 & 0.22 & 0.30 & -0.31 & 0.00 & 1.00 & -0.31 & -0.49 & 0.25 & -0.35 & 0.20 & 0.31 & -0.15 & -0.25 & 0.79 \\ \hline
 $a_0^{T_2}$ & 1.00 & -0.96 & 0.16 & 1.00 & 0.98 & -0.99 & 1.00 & -0.98 & -0.55 & 1.00 & -0.95 & -0.31 & 1.00 & 0.98 & -0.99 & 1.00 & -0.99 & 0.42 & 0.97 & -0.81 & 0.18 \\ \hline
 $a_1^{T_2}$ & 0.98 & -0.88 & -0.03 & 0.98 & 1.00 & -0.96 & 0.99 & -0.95 & -0.57 & 0.98 & -0.87 & -0.49 & 0.98 & 1.00 & -0.96 & 0.99 & -0.95 & 0.33 & 0.93 & -0.70 & 0.01 \\ \hline
 $a_2^{T_2}$ & -0.99 & 0.96 & -0.21 & -0.99 & -0.95 & 0.99 & -0.98 & 0.99 & 0.49 & -0.99 & 0.96 & 0.25 & -0.99 & -0.96 & 1.00 & -0.98 & 0.98 & -0.45 & -0.97 & 0.83 & -0.20 \\ \hline
 $a_0^{T_3}$ & 1.00 & -0.94 & 0.12 & 1.00 & 0.99 & -0.99 & 1.00 & -0.98 & -0.53 & 1.00 & -0.94 & -0.35 & 1.00 & 0.99 & -0.98 & 1.00 & -0.98 & 0.43 & 0.96 & -0.79 & 0.14 \\ \hline
 $a_1^{T_3}$ & -0.99 & 0.98 & -0.26 & -0.99 & -0.94 & 0.99 & -0.98 & 1.00 & 0.44 & -0.99 & 0.97 & 0.20 & -0.99 & -0.95 & 0.98 & -0.98 & 1.00 & -0.56 & -0.97 & 0.83 & -0.24 \\ \hline
 $a_2^{T_3}$ & 0.43 & -0.55 & 0.54 & 0.42 & 0.33 & -0.48 & 0.42 & -0.55 & 0.35 & 0.42 & -0.55 & 0.31 & 0.42 & 0.33 & -0.45 & 0.43 & -0.56 & 1.00 & 0.41 & -0.45 & 0.27 \\ \hline
$a_0^{A_0}$ & 0.98 & -0.98 & 0.31 & 0.98 & 0.92 & -0.98 & 0.96 & -0.95 & -0.61 & 0.97 & -0.98 & -0.15 & 0.97 & 0.93 & -0.97 & 0.96 & -0.97 & 0.41 & 1.00 & -0.91 & 0.35 \\ \hline
$a_1^{A_0}$ & -0.81 & 0.93 & -0.65 & -0.81 & -0.69 & 0.83 & -0.78 & 0.81 & 0.55 & -0.81 & 0.94 & -0.25 & -0.81 & -0.70 & 0.83 & -0.79 & 0.83 & -0.45 & -0.91 & 1.00 & -0.71 \\ \hline
$a_2^{A_0}$ & 0.18 & -0.43 & 0.91 & 0.18 & 0.00 & -0.21 & 0.14 & -0.20 & -0.26 & 0.18 & -0.45 & 0.79 & 0.18 & 0.01 & -0.20 & 0.14 & -0.24 & 0.27 & 0.35 & -0.71 & 1.00 
\\ \hline \hline
            \end{tabular} }
        \caption{Correlation between the fitted coefficients of $z$-expansion for the local form factors.}
    \label{tab:correlation}
\end{table}

\section{Transversity amplitudes and angular coefficients}\label{app:tAmp}

The non-local corrections are included through a shift in the Wilson coefficient $C_9^\text{eff.}$ in the transversity amplitudes. We are using the following definitions for the transversity amplitudes from Ref.~\cite{Altmannshofer:2008dz}
\begin{align}
A_\perp^{L,R}  &=  N \sqrt{2} \tilde{\lambda}_{q^2}^{1/2} \bigg[ 
\left[ (C_9^\text{eff.}+\Delta C_{9,\perp} )  \mp C_{10}^\text{eff.}  \right] \frac{ V(q^2) }{ m_{B_s}  + m_\phi} + \frac{2m_b}{q^2} C_7^\text{eff.}  T_1(q^2)
\bigg],\\
A_\parallel^{ L,R}  & = - N \sqrt{2}(m_{B_s} ^2 - m_\phi^2) \bigg[ \left[ (C_9^\text{eff.}+\Delta C_{9,\parallel} ) \mp C_{10}^\text{eff.}  \right] 
\frac{A_1(q^2)}{m_{B_s} -m_\phi} +\frac{2 m_b}{q^2} C_7^\text{eff.}  T_2(q^2)
\bigg],\\
A_0^{L,R}  &=  - \frac{N}{2 m_\phi \sqrt{q^2}}  \bigg\{ 
 \left[ (C_9^\text{eff.}+\Delta C_{9,0}) \mp C_{10}^\text{eff.}  \right]
\bigg[ (m_{B_s} ^2 - m_\phi^2 - q^2) ( m_{B_s}  + m_\phi) A_1(q^2) 
 - \frac{\tilde{\lambda}_{q^2} A_2(q^2)}{m_{B_s}  + m_\phi}
\bigg] 
\nonumber\\
& \qquad + {2 m_b} C_7^\text{eff.}  \bigg[
 (m_{B_s} ^2 + 3 m_\phi^2 - q^2) T_2(q^2)
-\frac{\tilde{\lambda}_{q^2}}{m_{B_s} ^2 - m_\phi^2} T_3(q^2) \bigg]
\bigg\}, \\
 A_t  &= \frac{2 N}{\sqrt{q^2}}\tilde{\lambda}_{q^2}^{1/2}   C_{10}^\text{eff.}  A_0(q^2) .
\end{align}
The quantity 
\begin{equation}
N \equiv \vert V_{tb}^{*}V_{ts}^*\vert \left[\frac{G_F^2 \alpha_{\rm EM}^2}{3\cdot 2^{10}\pi^5 m_{B_s} ^3}
 q^2 \tilde{\lambda}_{q^2}^{1/2}
\beta_\mu \right]^{1/2},
\end{equation}
is a normalization factor and $\beta_\mu=\sqrt{1-4m_\mu^2/q^2}$ where $m_\mu=105.6\MeV$ is the muon mass. 

The differential decay width and the $CP$-averaged angular observables for $B_s \to \phi \mu^+\mu^-$ given in Eqs.~\eqref{eq:dgamma}
and \eqref{eq:angular_obs} are related to the angular coefficients expressed in terms of transversity amplitudes as~\cite{Altmannshofer:2008dz}
\begin{align}
  I_1^s & = \frac{(2+\beta_\mu^2)}{4} \left[|A_\perp^L|^2 + |A_\parallel^L|^2 + (L\to R) \right]  + \frac{4 m_\mu^2}{q^2} \Re\left(A_\perp^L A_\perp^{R^*} + A_\parallel^L A_\parallel^{R^*}\right), \\
  I_1^c & =  |A_0^L|^2 +|A_0^R|^2  + \frac{4m_\mu^2}{q^2} 
               \left[|A_t|^2 + 2\Re(A_0^L A_0^{R^*}) \right]  ,\\
  I_2^s & = \frac{ \beta_\mu^2}{4}\left[ |A_\perp^L|^2+ |A_\parallel^L|^2 + (L\to R)\right],\\
  I_2^c & = - \beta_\mu^2\left[|A_0^L|^2 + (L\to R)\right],\\
   I_3 & = \frac{ \beta_\mu^2}{2}\left[ |A_\perp^L|^2 -|A_\parallel^L|^2 + (L\to R)\right],\\
   I_4 & = \frac{ \beta_\mu^2}{\sqrt{2}}\left[\Re(A_0^L A_{||}^{L^*}) + (L\to R) \right] , \\
 I_7 & = \sqrt{2}\beta_\mu \left[\Im(A_0^L A_{||}^{L^*}) - (L\to R)  \right]\,. \label{eq:I7}
\end{align}

\bibliographystyle{bibstyle}
\bibliography{biblio.bib}

\end{document}